\documentclass[iicol, referee, sn-basic]{sn-jnl}

\usepackage{graphicx}%
\usepackage{multirow}%
\usepackage{amsmath,amssymb,amsfonts}%
\usepackage{amsthm}%
\usepackage{mathrsfs}%
\usepackage[title]{appendix}%
\usepackage{xcolor}%
\usepackage{textcomp}%
\usepackage{manyfoot}%
\usepackage{booktabs}%
\usepackage{algorithm}%
\usepackage{algorithmicx}%
\usepackage{algpseudocode}%
\usepackage{listings}%
\usepackage{subcaption}

\theoremstyle{thmstyleone}%
\theoremstyle{thmstyletwo}%

\theoremstyle{thmstylethree}%

\begin{document}

\title[Article Title]{Validation of satellite and reanalysis rainfall products against rain gauge observations in Ghana and Zambia}




\author*[1,2]{\fnm{John} \sur{Bagiliko}}\email{	john.bagiliko@aims-senegal.org}

\author[3]{\fnm{David} \sur{Stern}}

\author[1]{\fnm{Denis} \sur{Ndanguza}}

\author[4]{\fnm{Francis Feehi} \sur{Torgbor}}

\affil*[1]{\orgdiv{Department of Mathematics}, \orgname{School of Science, College of Science and Technology, University of Rwanda}, \orgaddress{\city{Kigali}, \postcode{P.O. Box 3900}, \country{Rwanda}}}

\affil[2]{ \orgname{African Institute for Mathematical Sciences, Research and Innovation Center}, \orgaddress{\street{Rue KG590 ST}, \city{Kigali}, \country{Rwanda}}}

\affil[3]{ \orgname{IDEMS International}, \orgaddress{\street{RG2 7AX}, \city{Reading}, \country{United Kingdom}}}

\affil[4]{ \orgname{Ghana Innovations in Development, Education and the Mathematical Sciences}, \orgaddress{\street{Okponglo, East Legon}, \city{Accra}, \country{Ghana}}}


\abstract{
	Accurate rainfall data are crucial for effective climate services, especially in Sub-Saharan Africa, where agriculture depends heavily on rain-fed systems. The sparse distribution of rain-gauge networks necessitates reliance on satellite and reanalysis rainfall products (REs). This study evaluated eight REs---CHIRPS, TAMSAT, CHIRP, ENACTS, ERA5, AgERA5, PERSIANN-CDR, and PERSIANN-CCS-CDR---in Zambia and Ghana using a point-to-pixel validation approach. The analysis covered spatial consistency, annual rainfall summaries, seasonal patterns, and rainfall intensity detection across 38 ground stations. Results showed no single product performed optimally across all contexts, highlighting the need for application-specific recommendations. All products exhibited a high probability of detection (POD) for dry days in Zambia and northern Ghana (70\% $<$ POD $<$ 100\%, and 60\% $<$ POD $<$ 85\%, respectively), suggesting their utility for drought-related studies. However, all products showed limited skill in detecting heavy and violent rains (POD close to 0\%), making them unsuitable for analyzing such events (e.g., floods) in their current form. Products integrated with station data (ENACTS, CHIRPS, and TAMSAT) outperformed others in many contexts, emphasizing the importance of local observation calibration. Bias correction is strongly recommended due to varying bias levels across rainfall summaries. A critical area for improvement is the detection of heavy and violent rains, with which REs currently struggle. Future research should focus on this aspect.
}




\maketitle

\section{Introduction}\label{sec1}

Accurate historical rainfall observations are critical for providing reliable climate services \citep{Ageet2022, amt-11-1921-2018, Dinku2018, duPlessis2021}, particularly in regions where agriculture and water resources are highly sensitive to climatic variations \citep{MANTON2010184}. In Sub-Saharan Africa, where approximately 95\% of agriculture is rain-fed \citep{abrams2018unlocking, Nyoni2024}, access to precise rainfall data is indispensable for effective agricultural planning and management. For instance, recent floods and droughts have led to poor crop yields or complete crop losses across the continent \cite{IPCC2022} including Ghana and Zambia, underscoring the need for reliable rainfall information in these places. However, weather station and rain-gauge networks in this region are often sparse, both spatially and temporally, or in some cases, entirely non-existent \citep{Li2013, Ageet2022, Cocking2024}. This scarcity of ground-based observations poses significant challenges for generating localized climate information, further complicating efforts to mitigate the impacts of climate variability on agricultural productivity and food security.

REs offer a promising solution by providing rainfall estimates with high spatial and temporal resolutions \citep{Feidas2009, Monsieurs2018}. These REs can potentially enhance climate monitoring and forecasting capabilities in regions with limited ground-based observation networks. However, the REs do not come without potential shortcomings. They are prone to uncertainties arising from indirect estimation methods, reliance on proxy variables, and potential errors in temporal sampling, course resolution, algorithms, or sensor accuracy, which can limit their reliability for climate applications \citep{amt-11-1921-2018, Dinku2018, Tot2015}. Therefore their accuracy and usability must be rigorously validated against available ground-based observations \citep{Dinku2018, amt-11-1921-2018, Ageet2022} before they can be effectively integrated into climate services. Validation serves as a key step in assessing how closely these estimates align with actual observations, ensuring their reliability for applications such as agriculture, hydrology, and disaster management \citep{Feidas2009}.

Numerous validation studies have been carried out across the African continent, including those by \cite{DOSSANTOS2022101168, Dinku2007, Dinku2018, Kimani2017, Ageet2022, Feidas2009, Monsieurs2018, Tarnavsky2014, Maranan2020, Hofstra2009a, Hofstra2009b, amt-11-1921-2018, Tot2015, Garba2023, Young14, Katsekpor2024, Mekonnen2023, Gebremicael2019, MAPHUGWI2024107718} and \cite{GASHAW2023100994}. These studies commonly utilize two main approaches for evaluation: pixel-to-pixel and point-to-pixel comparisons \citep{Saemian2021}. Each approach has distinct advantages and limitations, depending on the specific application and the availability of ground-based observational data. 

Pixel-to-pixel validation involves gridding a dense network of rain-gauge stations to a resolution similar to that of the satellite or reanalysis product, allowing for direct comparison across equivalent spatial scales \citep{Saemian2021, Dinku2007, Kimani2017}. This method is particularly useful for obtaining a broad, regional assessment of the product's performance. It is well-suited for applications that require generalized climate information, such as drought monitoring or regional climate modeling. A key advantage of pixel-to-pixel validation is its ability to provide a large-scale comparison that captures spatial rainfall patterns over a wide area. For instance, \cite{Dinku2007} conducted a pixel-to-pixel validation in Ethiopia using multiple REs, revealing insights into how they perform over various climatic regions. \cite{Kimani2017} also used pixel-to-pixel validation in East Africa to evaluate the performance of REs, and concluded that while the REs exhibited systematic underestimations, particularly in orographic regions during the October to December rainy season, they could replicate rainfall patterns. One drawback of pixel-to-pixel validation is that it may obscure localized variations, as the averaging effect within grid cells can mask important small-scale rainfall events that are critical for localized agricultural applications.

In contrast, point-to-pixel validation \citep{Maranan2020, Ageet2022, Monsieurs2018, Dinku2018, Saemian2021, amt-11-1921-2018} directly compares REs with point-based rain-gauge observations. One advantage of point-to-pixel validation is its ability to provide a more precise assessment at specific locations, helping to identify discrepancies between satellite estimates and on-the-ground observations.  However, a key constraint of this approach is the fact that gauges give point measurements while satellites and reanalysis produce spatial averages, which sometimes struggle with local convective storms and orographic rainfall \citep{Monsieurs2018}. Comparing rainfall measured at a specific point to satellite estimates averaged over a large area is inherently problematic \citep{Monsieurs2018}.

While we acknowledge the limitations of the point-to-pixel approach, this study conducted a comprehensive validation of eight REs in Zambia and Ghana using this method. This was chosen primarily because we sought to evaluate the suitability of these REs for providing localized climate information. This is due to the growing demand for gridded data with finer spatial and temporal resolutions, driven by the shift in climate change research from global analyses to more regional and localized investigations \citep{Hofstra2009a}. Additionally, the available station network was sparse, with no more than one station located within a given satellite or reanalysis grid. The uncertainty of interpolation methods increases significantly as the number of stations within a grid decreases \citep{Hofstra2009a, Hofstra2009b, Maidment2017}. The native resolutions of the REs were retained (as also done by \cite{hess-21-6201-2017}, and \cite{amt-11-1921-2018}) without remapping them to a common resolution. This decision was intentional, as it allowed us to assess the performance of each RE as they are in their operational forms. While remapping to a common resolution could reduce spatial discrepancies, it might also obscure the inherent characteristics of each RE. Zambia and Ghana were selected because they represent distinct yet complementary rainfall regimes --- unimodal in Zambia and northern Ghana, and bimodal in southern Ghana (see Figures~\ref{fig2_a_b_} and \ref{fig_3}) --- while also sharing similarities in their reliance on rain-fed agriculture and challenges related to sparse station networks. This combination allowed for a robust evaluation of REs across diverse climatic and geographical contexts.

Our validation focused on key aspects, including spatial consistency, annual rainfall summaries, seasonal patterns, and rainfall intensity detection. The importance of evaluating REs under diverse conditions has been emphasized in previous studies \citep{Dinku2018}. Through this multi-dimensional approach, we aimed to provide a robust assessment of the utility of these REs for climate services. Besides, this study evaluated a wide range of products, including the Precipitation Estimation from Remotely Sensed Information using Artificial Neural Networks-Cloud Classification System-Climate Data Record (PERSIANN-CCS-CDR; hereafter PCCSCDR) \citep{Sadeghi2021} and the Precipitation Estimation from Remotely Sensed Information using Artificial Neural Networks-Climate Data Record (PERSIANN-CDR; hereafter PCDR) \citep{Ashouri2015}, which have not been extensively validated across the African continent, especially at our study area. By including these products, we provide new insights to the scientific community. Furthermore, the spatial consistency analysis conducted in this study, an aspect often overlooked in validation efforts, offers an understanding of its significance in assessing the reliability of REs.

The remainder of this paper is structured as follows: Section \ref{sec_method} describes the materials and methods used for the validation, while the results are presented in Section \ref{section_results}, and discussed in Section \ref{section_discussion}. Finally, the conclusion is given in Section \ref{sec_conclusion}. The terms "gauge observations", or "station data", are used interchangeably to refer to rain-gauge station observations or summaries derived from rain-gauge station observations.

\section{Materials and Methods}\label{sec_method}
\subsection{Study Area}

\begin{figure*}[ht]%
	\centering
	\includegraphics[width=1\textwidth]{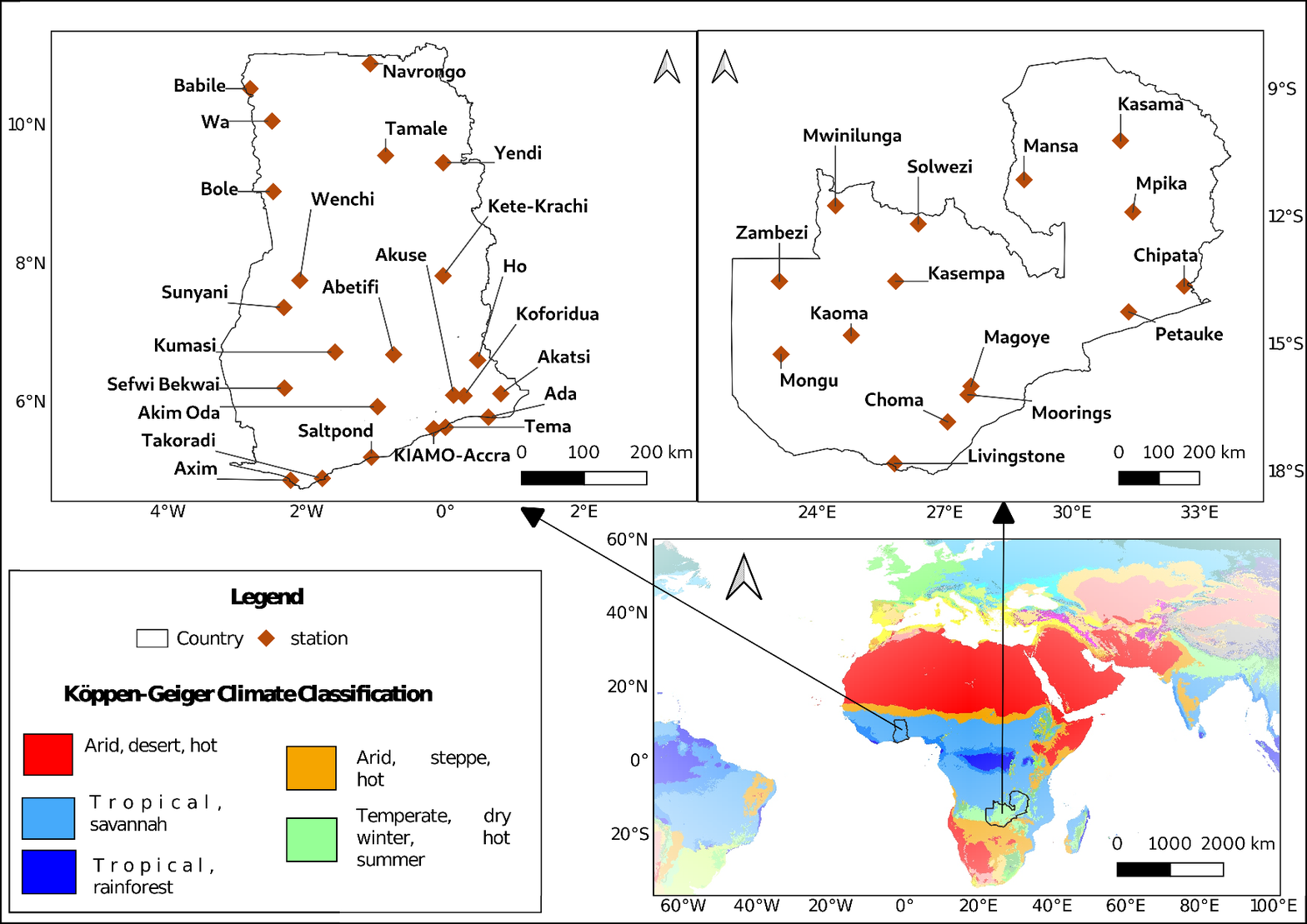}
	\caption{Map of study area showing the station locations in Ghana (at the top left) and Zambia (at the top right) and the dominant climatological zones in Africa (at the bottom right) based on K\"oppen-Geiger climate classification \citep{Beck2023} --- (1991-2020)}\label{fig1}
\end{figure*}

The validation area covers Zambia and Ghana (Figures~\ref{fig1}). Zambia lies between 22$^{\circ}$ to 34$^{\circ}$ \citep{jain2007empirical} east of Greenwich, and 8$^{\circ}$ to 18$^{\circ}$ south of the equator \citep{Kaczan2013}. The prevailing climate is mainly sub-tropical, witnessing around 95 percent of its total precipitation during the wet season extending from October to April \citep{Maidment2017} when the Inter-tropical Convergence Zone (ITCZ) is located in the region \citep{hachigonta2008analysis}. Zambia is a landlocked, relatively flat country where rainfall is predominantly the result of convection \citep{Maidment2017}. The rainfall in this region tends to be more uniform and widespread. The variations in rainfall tend to be tied to the specific agroecological region. In the northern areas, the average annual precipitation exceeds 1200 mm/year, while the southern part experiences less than 700 mm/year annually \citep{Kaczan2013}. The central part receives an annual rainfall of approximately 800 - 1000 mm/year, distributed evenly across the crop growing season  \citep{jain2007empirical}.   

Ghana is located in West Africa, and lies between latitudes 4.6$^{\circ}$N and 11$^{\circ}$N, and longitudes -3.3$^{\circ}$W and 1.8$^{\circ}$E \citep{Oduro2024}. The climate of Ghana is tropical and humid \citep{boateng2021rainfall}. Two rainy seasons occur in the south from April to July and from September to November (with peaks in June and October respectively), whereas the north has only one rainy season from April to September, with one peak around August and September \citep{Amekudzi2015-qk}. Unlike Zambia, the rainfall pattern in Ghana is quite complex with coastal and orographic influences. The seasonality in the rainfall patterns is brought about by the movement of the Inter-Tropical Convergence Zone (ITCZ) \citep{boateng2021rainfall, torgbor2018rainfall}. The country can be broadly categorized under three climatic zones, which are the Savannah zone, the Forest zone, and the Coastal zone \citep{Bessah2022, Oduro2024}. Mean annual rainfall ranges between 900  and 2100 mm, where the south-west has relatively high values while the north, and south-eastern coast have relatively low values \citep{Atiah2020-rz}. 

\begin{figure*}[ht]
	\subfloat[]{
		\begin{minipage}[1\width]{
				0.5\textwidth}
			\centering
			\includegraphics[width=1\textwidth]{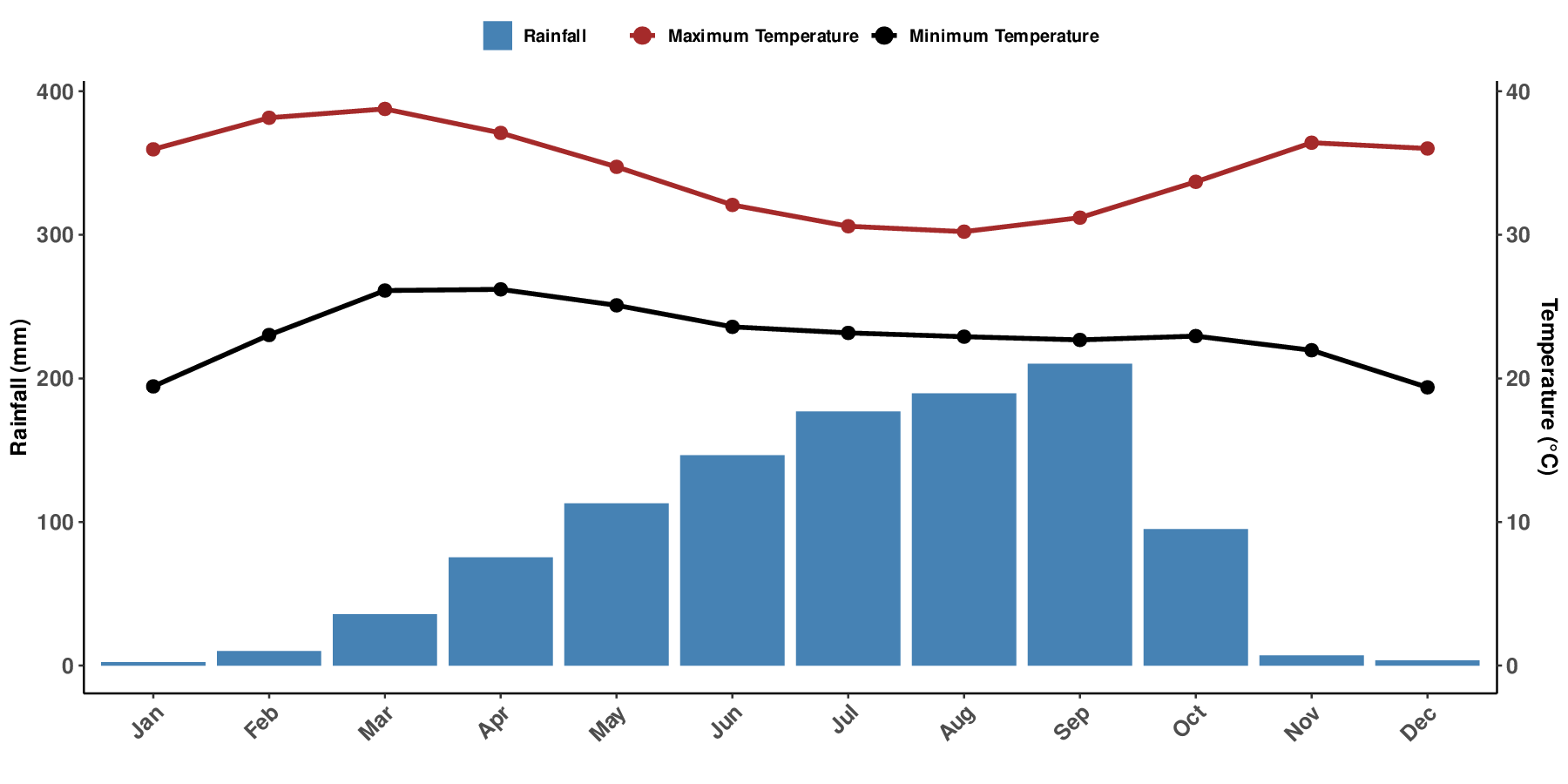}
	\end{minipage}}
	\hfill 	
	\subfloat[]{
		\begin{minipage}[1\width]{
				0.5\textwidth}
			\centering
			\includegraphics[width=1\textwidth]{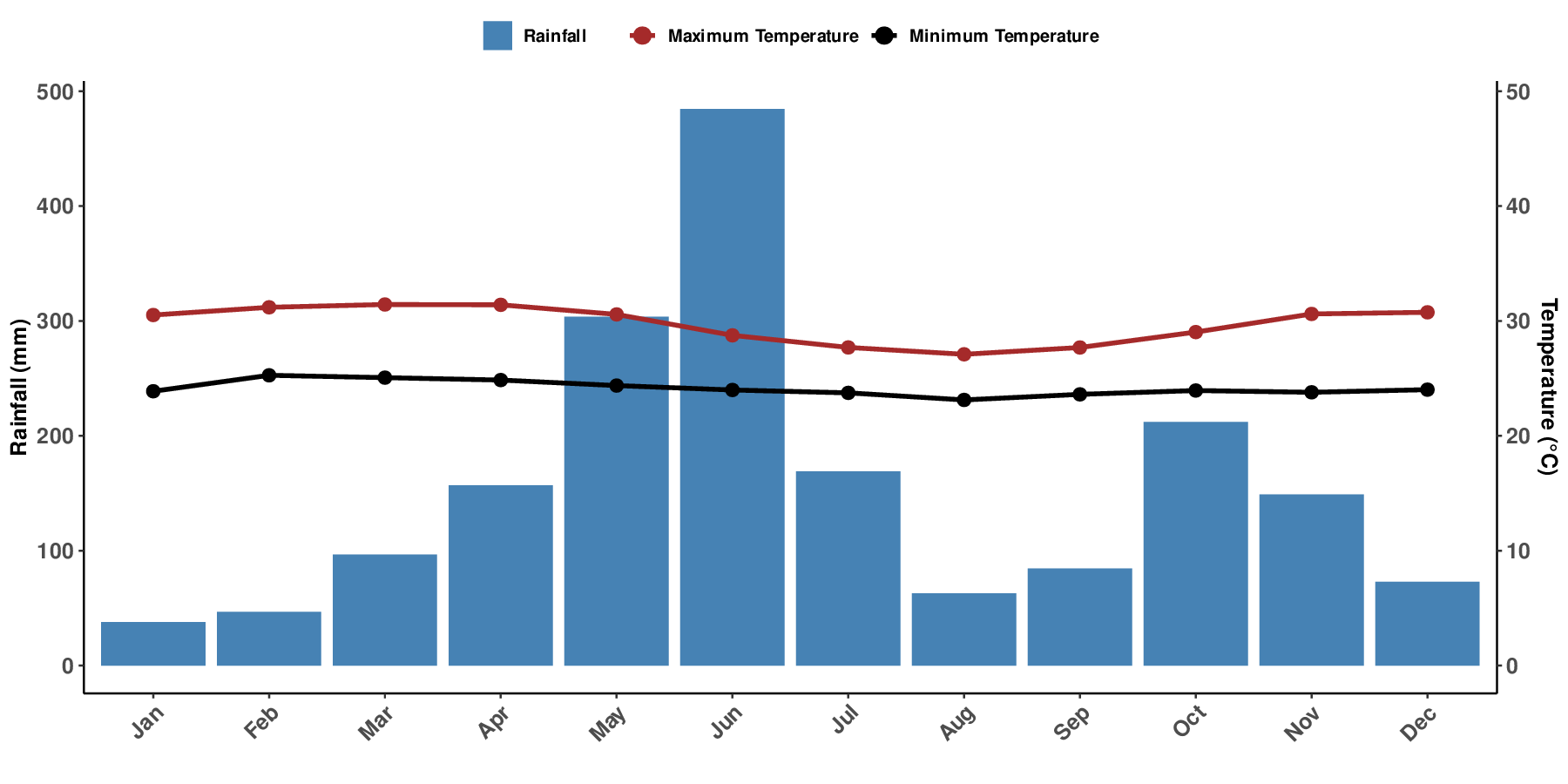}
	\end{minipage}}
	\caption{Ombrothermic diagrams for Tamale (a), Axim (b), chosen to represent the unimodal and bimodal rainfall patterns in the northern and southern parts of Ghana, respectively. The bar charts show the mean monthly total rainfalls while the line plots show the mean maximum temperatures (red) and mean minimum temperatures (black). The Temperature scale $=$ Rainfall scale / 10. Ombrothermic diagrams for other stations are in the Supplementary Materials}\label{fig2_a_b_}
\end{figure*}

\begin{figure}[ht]%
	\centering
	\includegraphics[width=0.5\textwidth]{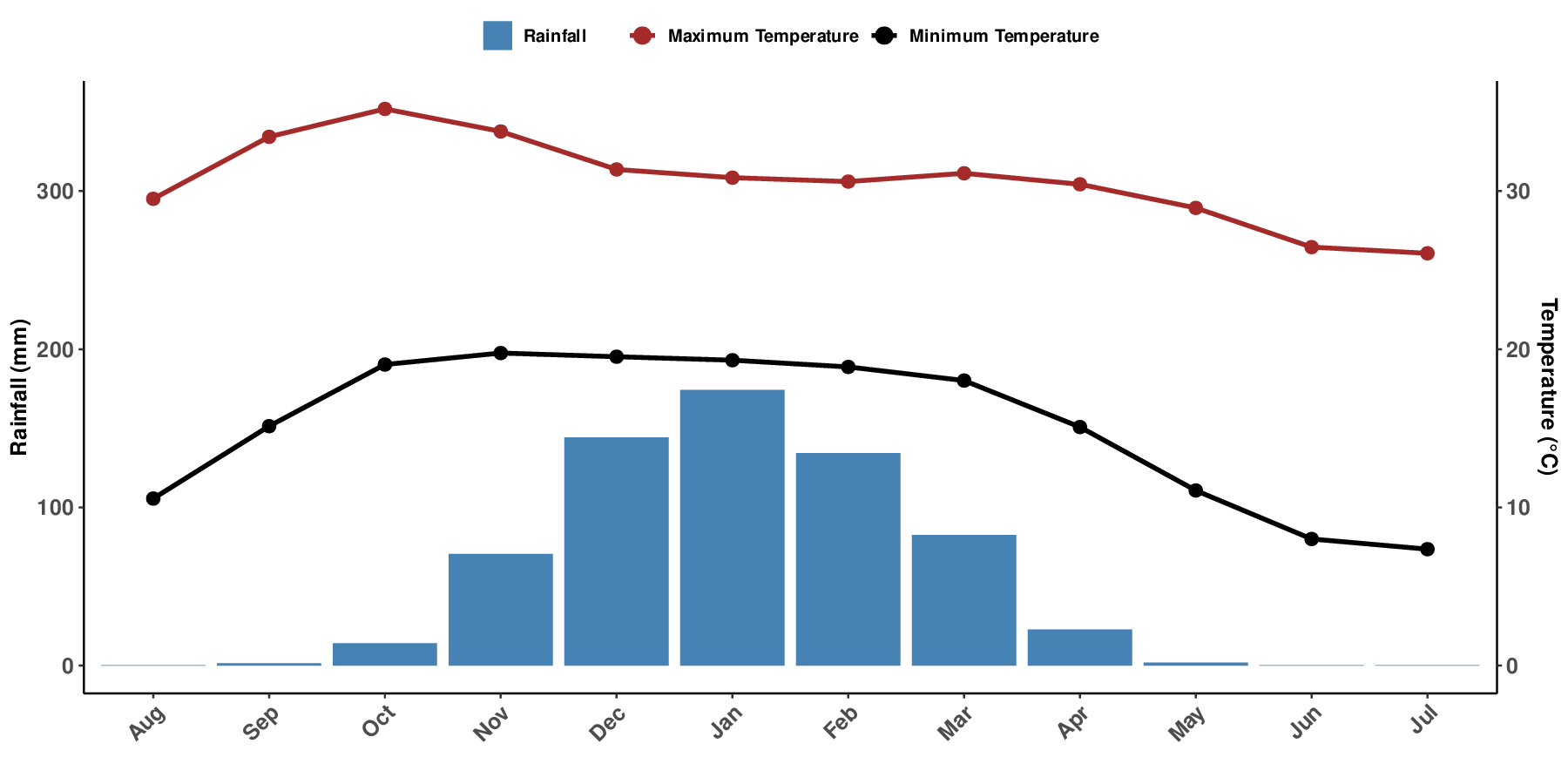}
	\caption{Ombrothermic diagram for Livingstone showing the unimodal rainfall pattern in the in Zambia (just as the other stations in the Supplementary Material). The bar charts show the mean monthly total rainfalls while the line plots show the mean maximum temperatures (red) and mean minimum temperature (black). The Temperature scale $=$ Rainfall scale / 10. The year at this station (and all Zambian stations in this work) have been considered to start in August and end in July)}\label{fig_3}
\end{figure}

The study area was selected based on two key factors:

The first factor was the objective to capture three distinct rainfall patterns across Africa:

\begin{enumerate} 
	\item A unimodal rainfall pattern in the northern part of Sub-Saharan Africa, observed in places such as northern Ghana (as shown on Figure~\ref{fig2_a_b_}a), Burkina Faso, Mali and Niger 
	\item A unimodal rainfall pattern in the southern part of the continent, characteristic of regions like Zambia (Figure~\ref{fig_3}), Mozambique, and Malawi. 
	\item A bimodal rainfall pattern in the mid-belt of the continent, found in areas including southern Ghana (Figure~\ref{fig2_a_b_}b) and parts of Kenya.
 \end{enumerate}
By evaluating the performance of REs across these varied rainfall regimes, the study aimed to provide insights into their potential performance in other regions with similar climatic patterns. However, these results are not to be generalized to other locations without validation.

The second factor was the availability of observed rainfall data. We are grateful to the Ghana Meteorological Agency, and the Zambia Meteorological Department for their collaboration and for providing station data for this study to be carried out.

\subsection{Data}
\subsubsection{Station Data}
Data from $15$ stations across the country of Zambia, and $23$ stations across Ghana were used in this validation study. The Zambian data was obtained from the Zambia Meteorological Department (ZMD), while the Ghanaian data was obtained from the Ghana Meteorological Agency (GMet). They were daily rain gauge measurements. The data, as were obtained from ZMD and GMet, had longer records going as far back as 1930 for some of the stations. However, 1983 was chosen as the starting point for the stations to align with the years for which most of the REs have available data. Zambian stations with long records (at least 30 years, with the exception of two stations which had 27 and 28 years) were used, and they are scattered unevenly across all the three climatic zones of the country. In the case of Ghana, the 23 stations are synoptic stations \citep{Oduro2024} and are distributed across the three climatic zones. 

A key challenge associated with station data is its potential for errors or inconsistencies. To address this, all station observations underwent rigorous quality-control procedures before being utilized in the study. These procedures included checks for consecutive rain days, repeated identical rainfall values, extreme values, dry months, and missing data. This process was conducted in close collaboration with the data providers and required significant time and effort, given its critical role in ensuring data reliability. Values that failed the quality-control checks were flagged as missing. Only stations with at least 70\% non-missing data were included in the analysis. For annual summaries, years with fewer than 355 days of non-missing observations were excluded. Similarly, for seasonal and categorical comparisons, only days with non-missing observations in both the station data and the REs were considered. Table \ref{tab_stdata} provides information on the stations used, including their names, geographic coordinates, temporal coverage, and the percentage of complete data within the study period.
\begin{table*}[h]
	\caption{Details of 38 stations in Zambia and Ghana considered for the study}\label{tab_stdata}%
	\begin{tabular}{@{}lllllll@{}}
		\toprule
		Country & Station	& Latitude & Longitude & Elevation (m)	& Period & Complete Days (\%)\\
		\midrule
		Zambia & Chipata &	-13.64 &	32.64 & 1025 &	1983 - 2022 & $98.2$ \\
		Zambia & Kasama &	-10.22 &	31.14 & 1384 &	1983 - 2018 & $83.7$\\
		Zambia & Mansa &	-11.14 &	28.87 & 1257 &	1983 - 2017 & $82.3$\\
		Zambia & Mpika &	-11.90 &	31.43 & 1399	& 1983 - 2020 & $83.7$ \\
		Zambia & Magoye &	-16.00 &	27.62 & 1025	& 1983 - 2016 & $81.4$\\
		Zambia & Moorings &	-16.15 &	27.32 & 1085	& 1983 - 2010 & $68.3$ \\
		Zambia & Choma &	-16.84 &	27.07 & 1275	& 1983 - 2011 & $71.2$\\
		Zambia & Livingstone &	-17.82 &	25.82 & 991	& 1983 - 2019 & $89.2$ \\
		Zambia & Petauke &	-14.25 &	31.33 & 1022	& 1983 - 2022 & $96.4$\\
		Zambia & Kaoma &	-14.80 &	24.80 & 1152 & 1983 - 2022 & $83.1$ \\
		Zambia & Kasempa &	-13.53 &	25.85 & 1134 & 1983 - 2022 & $72.5$ \\
		Zambia & Mongu &	-15.25 &	23.15 & 1053 & 1983 - 2021 & $95.8$ \\
		Zambia & Mwinilunga &	-11.75 &	24.43 & 1363 & 1983 - 2022 & $84.0$ \\
		Zambia & Solwezi &	-12.18 &	26.38 & 1333 & 1983 - 2022 & $99.0$ \\
		Zambia & Zambezi &	-13.53 &	23.11 & 1078 & 1983 - 2022 & $96.8$ \\
		Ghana & Abetifi & 6.680 & -0.747 & 594.7 & 1983 - 2022 & $92.7$  \\
		Ghana & Ada & 5.778 & 0.622 & 5.0 & 1983 - 2021 & 91.9  \\
		Ghana & Akatsi & 6.117 & 0.800 & 53.6 & 1983 - 2020 & 90.7  \\
	    Ghana & Akim Oda & 5.929 & -0.978 & 39.4 & 1983 - 2020 & $90.6$  \\
	    Ghana & Akuse & 6.095 & 0.119 & 17.4 & 1983 - 2020 & 94.6  \\
	    Ghana & Axim & 4.867 & -2.233 & 37.8 & 1983 - 2021 & 93.8 \\
	    Ghana & Babile & 10.517 & -2.817 & 304.7 & 1983 - 2022 & $94.8$  \\
	    Ghana & Bole & 9.033 & -2.483 & 299.5 & 1983 - 2022 & 98.1  \\
	    Ghana & Ho & 6.600 & 0.467 & 157.6 & 1983 - 2019 & 91.9  \\
	    Ghana & Kete-Krachi & 7.817 & -0.033 & 122.0 & 1983 - 2019 & $90.0$  \\
	    Ghana & KIAMO-Accra & 5.610 & -0.168 & 67.7 & 1983 - 2021 & $95.2$  \\
	    Ghana & Koforidua & 6.086 & 0.271 & 166.5 & 1983 - 2022 & $97.7$  \\
	    Ghana & Kumasi & 6.717 & -1.592 & 286.3 & 1983 - 2022 & $96.6$  \\
	    Ghana & Navrongo & 10.878 & -1.083 & 201.3 & 1983 - 2020 & $93.5$  \\
	    Ghana & Saltpond & 5.200 & -1.067 & 43.9 & 1983 - 2019 & $91.7$  \\
	    Ghana & Sefwi Bekwai & 6.197 & -2.321 & 170.8 & 1983 - 2021 & $96.0$  \\
	    Ghana & Sunyani & 7.359 & -2.330 & 308.8 & 1983 - 2022 & $98.3$  \\
	    Ghana & Takoradi & 4.894 & -1.774 & 4.6 & 1983 - 2022 & $97.3$  \\
	    Ghana & Tamale & 9.554 & -0.862 & 183.3 & 1983 - 2022 & $97.7$  \\
	    Ghana & Tema & 5.632 & 0.002 & 14.0 & 1983 - 2021 & 95.0  \\
	    Ghana & Wa & 10.05 & -2.50 & 322.7 & 1983 - 2020 & 90.4  \\	
		Ghana & Wenchi & 7.75 & -2.10 & 338.9 & 1983 - 2020 & 94.2  \\	
		Ghana & Yendi & 9.45 & -0.03 & 195.2 & 1983 - 2021 & 97.3  \\
		\botrule
	\end{tabular}
\end{table*}

\subsubsection{Satellite and Reanalysis Data}
The eight REs used in this study, which include six satellite-based and two reanalysis rainfall products, were the Climate Hazards Group Infrared Precipitation (CHIRP), CHIRP  with Stations (CHIRPS) \citep{Funk2014, Funk2015}, Tropical Application of Meteorology using Satellite data (TAMSAT) \citep{Maidment2017, Tarnavsky2014, Hersbach2020}, the European Centre for Medium-Range Weather Forecasts (ECMWF) v5 (ERA5) \citep{Hersbach2020, Bell2021}, fifth-generation reanalysis of ECMWF (AgERA5 hereafter AGERA5) \citep{boogaard2020agrometeorological, AgEra5_2020}, PCCSCDR \citep{Sadeghi2021}, PCDR \citep{Ashouri2015}, and Enhancing National Climate Services (ENACTS) \citep{Dinku2017, Dinku2022}. 

We selected these REs based on the following criteria:

\begin{itemize}
	\item Long Historical Records: Each product has a dataset extending over at least 30 years \citep{Funk2015, Maidment2017}, making them reliable for long-term climatic analysis.
	
	\item High Spatial and Temporal Resolution: The REs offer daily or sub-daily data at fine spatial resolutions, which is crucial for capturing localized rainfall patterns.
	
	\item REs which cover the study area, and provide data up to date. 
	
\end{itemize} 

In addition, we also wanted to cover a diverse range of REs  that represent different major approaches: 

\begin{itemize}
	\item Satellite-Based Products: CHIRP, CHIRPS, and TAMSAT  have undergone extensive validation in various African regions, and they have been used for a wide range of services on the continent \citep{Dinku2018}.
	
	\item Reanalysis Products: ERA5 \citep{Hersbach2020} is widely recognized as one of the top-performing reanalysis products, with a large body of studies supporting its use across Africa. AGERA5, derived from ERA5, relatively new, is tailored for agricultural and agro-ecological applications \citep{boogaard2020agrometeorological} but has not been extensively validated, particularly in Zambia and Ghana.
	
	\item Neural Network-Based Products: PCDR and PCCSCDR, both leveraging neural networks, have not been widely validated in the study area, even though PCDR has been recently validated in C\^ote d'Ivoire \citep{Kouakou2024}, parts of west and central Africa \citep{KOUAKOU2023101409}, Nigeria \citep {Ogbu2020}, and Northern Ghana \citep{Katsekpor2024}.
	
	\item A Met Office-Led Product: ENACTS is a relatively new product that is led by Met offices of different African countries including the Zambia Meteorological Department. A relatively dense network of station data are included in the product, and has a potential added value for Africa \citep{siebert2019evaluation}. 
\end{itemize}

Although other REs, such as the African Rainfall Climatology version 2 (ARC2) \citep{Novella2013}, the Climate Prediction Center Morphing Method (CMORPH) \citep{joyce2004cmorph}, Integrated Multi-satellitE Retrievals for GPM (IMERG) \citep{PRADHAN2022112754},  and the Tropical Rainfall Measuring Mission (TRMM) \citep{simpson1988proposed} have been validated in various studies, they failed to meet one or two of the above-mentioned criteria; hence, they were not included in our evaluation. \cite{Kumar2024} and \cite{Saemian2021} provide an extensive list of 23 and 44 well-known products respectively with relevant details.

We employed a point-to-pixel approach, extracting daily rainfall data from the REs at their grid points closest to the station coordinates for comparison with observational data. However, for certain REs (e.g., PCDR and TAMSAT), data for their nearest grid points to some stations were unavailable due to their location over the sea or largely outside the country's boundaries. In such cases, the affected stations were excluded from the validation process for those specific REs. Neither the station data were interpolated to grid scales nor were the REs remapped to a common resolution, as the rationale for this decision has been outlined in the introduction. A summary of the REs' datasets used in this study is provided in Table \ref{tab_satdata}.

\begin{sidewaystable}
	\caption{Details of satellite and reanalysis rainfall products}\label{tab_satdata}
	\begin{tabular*}{\textheight}{@{\extracolsep\fill}lccccc}
		\toprule%
		Product	& Inputs	& Spatial Coverage    & Period & Spatial Resolution  & Temporal Resolution\\
		\midrule
		CHIRPS & Satellite + gauge merge & Global & 1981 - present & 0.05$^{\circ}$ & Daily \\
		CHIRP & Satellite & Global & 1981 - present & 0.05$^{\circ}$ & Daily \\
		TAMSAT & Satellite + gauge calibration & Africa & 1983 - Present & 0.0375$^{\circ}$ & Daily  \\
		ERA5 & Reanalysis & Global & 1940 - present & 0.25$^{\circ}$ & Hourly\\
		AgERA5 & ERA5 & Global & 1979 - present & 0.1$^{\circ}$ & Daily\\
		ENACTS & Satellite + gauge merge & Selected countries (including Zambia) & 1981 (for Zambia) - present & 0.0375$^{\circ}$  &  Daily \\
		PCDR & Satellite & Global & 1983 - present & 0.25$^{\circ}$ & Daily\\
		PCCSCDR & Satellite & Global & 1983 - present & 0.04$^{\circ}$ & Daily\\
		\botrule
	\end{tabular*}
\end{sidewaystable}  

For REs that incorporate station data, such as CHIRPS, some studies exclude stations used in the product's development during the validation process to ensure an unbiased comparison \citep{Dinku2018}. While we acknowledge this approach as methodologically sound, in this study, we chose not to exclude any stations. Our rationale is twofold: first, we prioritized the perspective of end users, who are unlikely to know which stations are integrated into the products and simply seek guidance on which product is most suitable for their location and/or application of interest. Second, by including all stations in the validation, we gain insight into the product's performance both at locations used in its development and at those that were not. This approach allows us to assess the product's generalizability across unseen or arbitrary locations, providing a more comprehensive evaluation. While this was not the primary aim of our analysis, it is highlighted at specific points where it is particularly relevant. Third, many researchers compute summary statistics (e.g., mean bias, root mean squared error) across all available stations for each product to facilitate easy comparison and ranking. However, this approach can introduce bias if stations used in the product's development are included in the validation process, potentially inflating performance metrics \citep{Dinku2018, Tot2015}. Our study took a different approach; instead of focusing on the overall performance across all stations, we were interested in evaluating how each product performs at individual locations. While this method can be more labor-intensive, especially when working with a dense network of stations, it provides a more granular understanding of product accuracy and suitability for localized climate applications. This site-specific focus ensures that we assess each product's performance without the bias introduced by averaging across multiple stations.

 The ERA5 hourly values were accumulated to daily values. In the course of this study, PCCSCDR was no longer available for download and could not be downloaded for Ghana, and hence it was not considered in the validation for Ghana. Part of the reasons for that may be due to spatial consistency check in Section \ref{sec2} and Section \ref{sec3}. 

\subsection{Checking Spatial Consistency of REs}\label{sec2}
The spatial consistency of REs was examined, focusing on annual total rainfall, rain day frequency, and mean rain per rainy day. The interest was to ensure uniform spatial variation in these summaries, without station data comparison. Maps of the long-term averages of these summaries were used to study how well the climatology of Zambia and Ghana is captured by the REs, allowing for detection of any unnatural  spatial variations.

\subsection{Comparison of Annual Summaries}

Annual summaries (total rainfall, number of rainy days, and mean rain per rainy day) were used to evaluate the performance of the REs. In the case of Zambia, each year was adjusted to commence in August and end in July to capture the entire rain and dry seasons in one cycle.

These summaries were validated against observed rainfall data using a set of statistical metrics. The choice of these metrics was guided by their ability to capture different aspects of RE performance, ensuring a comprehensive evaluation. Specifically, the following metrics were employed:

\begin{itemize}
	\item \textbf{Mean Error (ME)} (\ref{eqn:me}): This metric quantifies the average difference between the REs and gauge observations. It provides a straightforward measure of overall bias, indicating whether the REs tend to overestimate (positive values) or underestimate (negative values) rainfall \cite{cli12110169}. This is particularly useful for identifying systematic errors in the REs. The ideal value for ME is 0, indicating no bias.
	
	\item \textbf{Percentage Bias (PBIAS)} (\ref{eqn:pbias}): PBIAS assesses the systematic bias of the REs relative to observed data, expressed as a percentage. It is valuable for understanding the magnitude and direction of bias \citep{Li2013}. PBIAS ranges from $-\infty$ to $+\infty$, with 0 indicating no bias, positive values indicating overestimation, and negative values indicating underestimation.
	
	\item \textbf{Linear Correlation Coefficient ($r$)} (\ref{eqn:r}): This metric evaluates the degree of linear relationship between the REs and gauge observations \citep{Yang2016}. It ranges from $-1$ to $+1$, where $+1$ indicates a perfect positive linear relationship, $-1$ indicates a perfect negative linear relationship, and 0 indicates no linear relationship. A high correlation indicates that the REs can reliably capture the temporal variability of rainfall.
	
	\item \textbf{Ratio of Standard Deviations (RSD)} (\ref{eqn:rsd}): RSD compares the variability of the REs with that of the observed rainfall data. An RSD of 1 indicates perfect agreement in variability, values greater than 1 indicate overestimation of variability, and values less than 1 indicate underestimation of variability. This metric is crucial for applications such as flood forecasting and drought monitoring.
\end{itemize}

These metrics are defined below:
\begin{equation}\label{eqn:me}
	\text{ME} = \frac{1}{N}\sum_{i=1}^{N} (S_{i} - O_{i})
\end{equation}

\begin{equation}\label{eqn:pbias}
	\text{PBIAS} = \frac{\sum_{i=1}^{N}(S_{i} - O_{i})}{\sum_{i=1}^{N}O_{i}} \times 100
\end{equation}

\begin{equation}\label{eqn:r}
	r =\frac{\sum_{i=1}^{N}({S}_{i}-\bar{S})({O}_{i}-\bar{O})}{\sqrt{\sum_{i=1}^{N}({S}_{i}-\bar{S})^{2}\sum_{i=1}^{N}({O}_{i}-\bar{O})^{2} }}
\end{equation}

\begin{equation}\label{eqn:rsd}
	\text{RSD} = \frac{{\sigma}_{s}}{{\sigma}_{o}}
\end{equation}

\noindent
where $S_{i}$ is the RE series, $O_{i}$ is the gauge observation series, $N$ is the number of data pairs, $\sigma_{s}$ and $\sigma_{o}$ are the standard deviations of the RE and gauge observation series respectively, $\bar{S}$ and $\bar{O}$ are the means of the RE and gauge observation series respectively.

\subsection{Comparison of Seasonal Behaviour}
To analyze rainfall frequency and intensity, we employed Markov chain models \citep{torgbor2018rainfall}, focusing on the Zero-Order Markov chain to estimate daily rainfall occurrence. The rainfall occurrence was modeled as a proportion for each day of the year using logistic regression, with Fourier series applied to capture the periodic nature of seasonal variation. This approach allowed for a detailed comparison between REs and gauge observations, facilitating an evaluation of rainfall frequency patterns across the annual cycle.

For each of the 15 stations in Zambia and the 23 stations in Ghana, separate zero-order Markov chain models incorporating three harmonics were fitted to both the gauge observations and the REs. The model formula is represented by (\ref{eq_season}) below: 

\begin{equation} 
	y(t) = \beta_{0} + \sum_{i=1}^{k} \Big[A_{i}\cos \left(\frac{2\pi it}{p}\right) + B_{i} \sin\left(\frac{2 \pi it}{ p}\right)\Big] + \epsilon \label{eq_season}, 
\end{equation}

\noindent where $k$ is the number of harmonics, $p$ represents the period, $t$ is the time (day of year), $\beta_{0}$, $A_{i}$, and $B_{i}$ are the model parameters, $\epsilon$ is the error term, and 

\begin{equation}
	y(t) = 
	\begin{cases} 
		0 & \text{if } x(t) < Tr \\
		1 & \text{Otherwise} 
	\end{cases} \label{eq:step_func1}
\end{equation}
\noindent
where $x(t)$ is the rainfall value on day $t$ in mm, and $Tr$ is the rainy day threshold in mm.

Importantly, only days where both gauge and product data were available were used in the models to ensure direct comparability.

Additionally, we explored the sensitivity of the results to varying rainy day thresholds. Separate models were developed for $Tr = $ 0.85 mm, 2 mm, 3 mm, 4 mm, and 5 mm for each product (while keeping the $Tr$ of the gauge at 0.85 mm) to examine whether adjusting rain day thresholds could mitigate biases. This approach enabled an assessment of how these biases and threshold adjustments might vary by location and time of year.

\subsection{Rainfall Intensity}\label{rain_intensity} 
In order to understand the ability of the REs to detect different types of rainfall, they were assessed over a set of rainfall intensity categories. The categories are defined in (\ref{eq:step_func}).  

The following terms were also used: 

\begin{itemize}
	\item True Positive (Hit): the number of
	observed rainfall events correctly detected by the RE.
	
	\item False Positive (False Alarm): the number of instances where the RE detected a rainfall event that was not observed.
	
	\item False Negative (Miss): the number of events where a rainfall event was observed but was not detected by the RE.
	
	\item True Negative: the number of instances where the RE correctly detects a dry day as observed.
\end{itemize}

\begin{equation}
	f(x) = 
	\begin{cases} 
		\text{Dry} & \text{if } x < 0.85 \\
		\text{Light Rain} & \text{if } 0.85 \leq x < 5 \\
		\text{Moderate Rain} & \text{if } 5 \leq x < 20 \\
		\text{Heavy Rain} & \text{if } 20 \leq x < 40 \\
		\text{Violent Rain} & \text{if } x \geq 40 
	\end{cases} \label{eq:step_func}
\end{equation}
\noindent
where $x$ is the daily rainfall value in mm. The intensity category definitions are similar to that of \cite{ZambranoBigiarini2017} except the dry day threshold is slightly lower based on reason by \cite{STERN2011}.

With rainfall intensity classification, we used POD, which is a standard measure for binary classification. It is the likelihood that the REs will correctly detect the same rainfall events on the same days as recorded in the gauge observations. The POD is given by (\ref{eqn:pod}).

\begin{equation}\label{eqn:pod}
	\text{POD} = \frac{\text{Hit}}{\text{Hit} + \text{Miss}}
\end{equation}

\section{Results}\label{section_results} 
The results from the study are presented under the following subsections:

\subsection{Spatial Consistency of REs} \label{sec3}
\begin{figure}[ht]%
	\centering
	\includegraphics[width=0.5\textwidth]{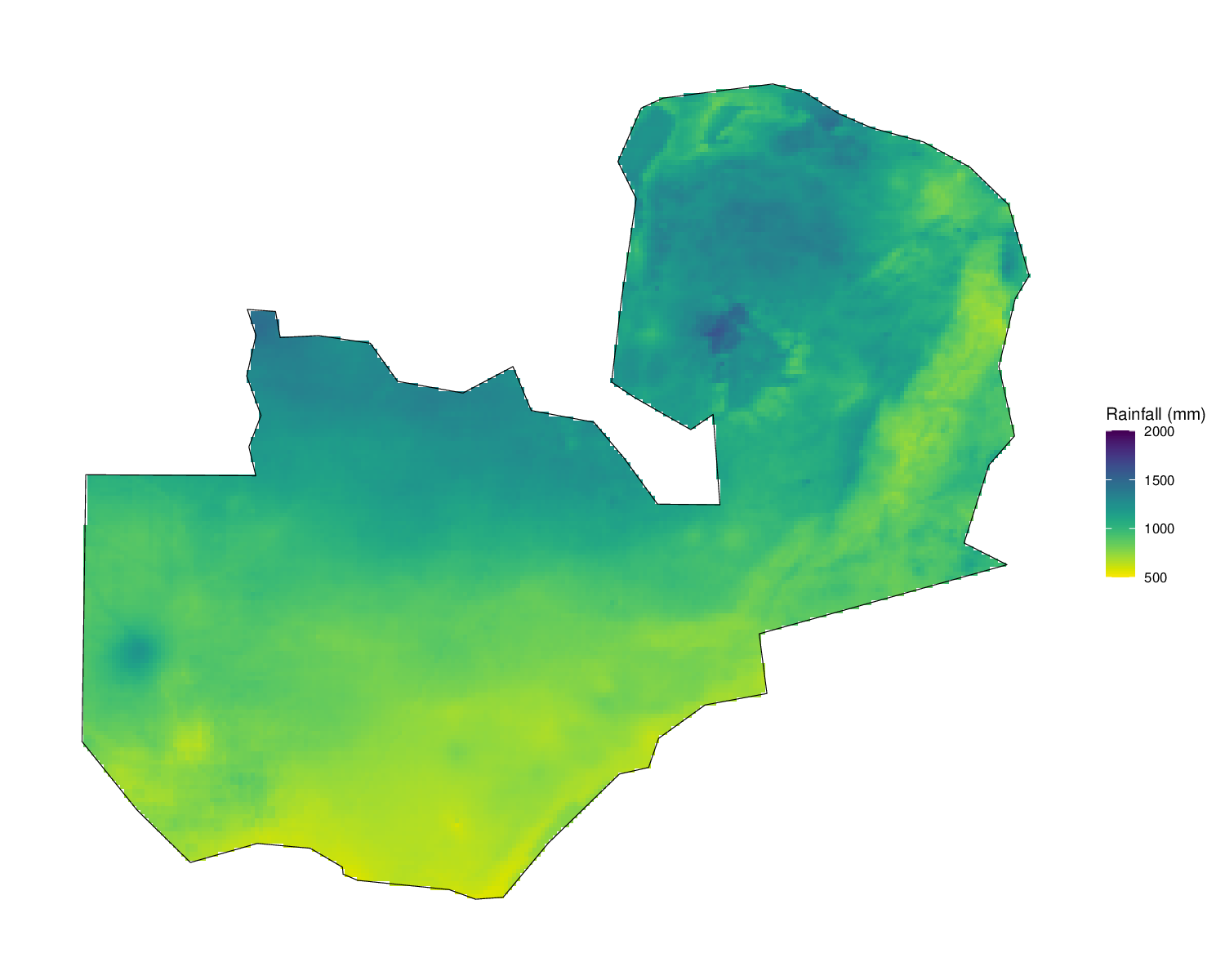}
	\caption{Mean total rainfall (mm/year) of CHIRPS for 1983-2022 in Zambia for checking the RE's spatial consistency}\label{fig4}
\end{figure}

For the case of Zambia, it was found that, with the exception of PCCSCDR, none of the REs exhibited any form of spatial irregularities for mean annual total rainfall, mean annual number of rainy days, and mean rain per rainy day. This is an indication that these REs most likely effectively capture Zambia's rainfall patterns. For instance, Figure \ref{fig4}, showing the spatial map of CHIRPS for mean annual total rainfall, reveals pronounced annual rainfall in the northern region (consistent with the climatology of Zambia), and lacking any noticeable unnatural variability. Consequently, rainfall estimates from nearby weather stations are likely to align closely. Applications relying on these summaries are likely to align well with gauge observations.  

\begin{figure}[ht]%
	\centering
	\includegraphics[width=0.5\textwidth]{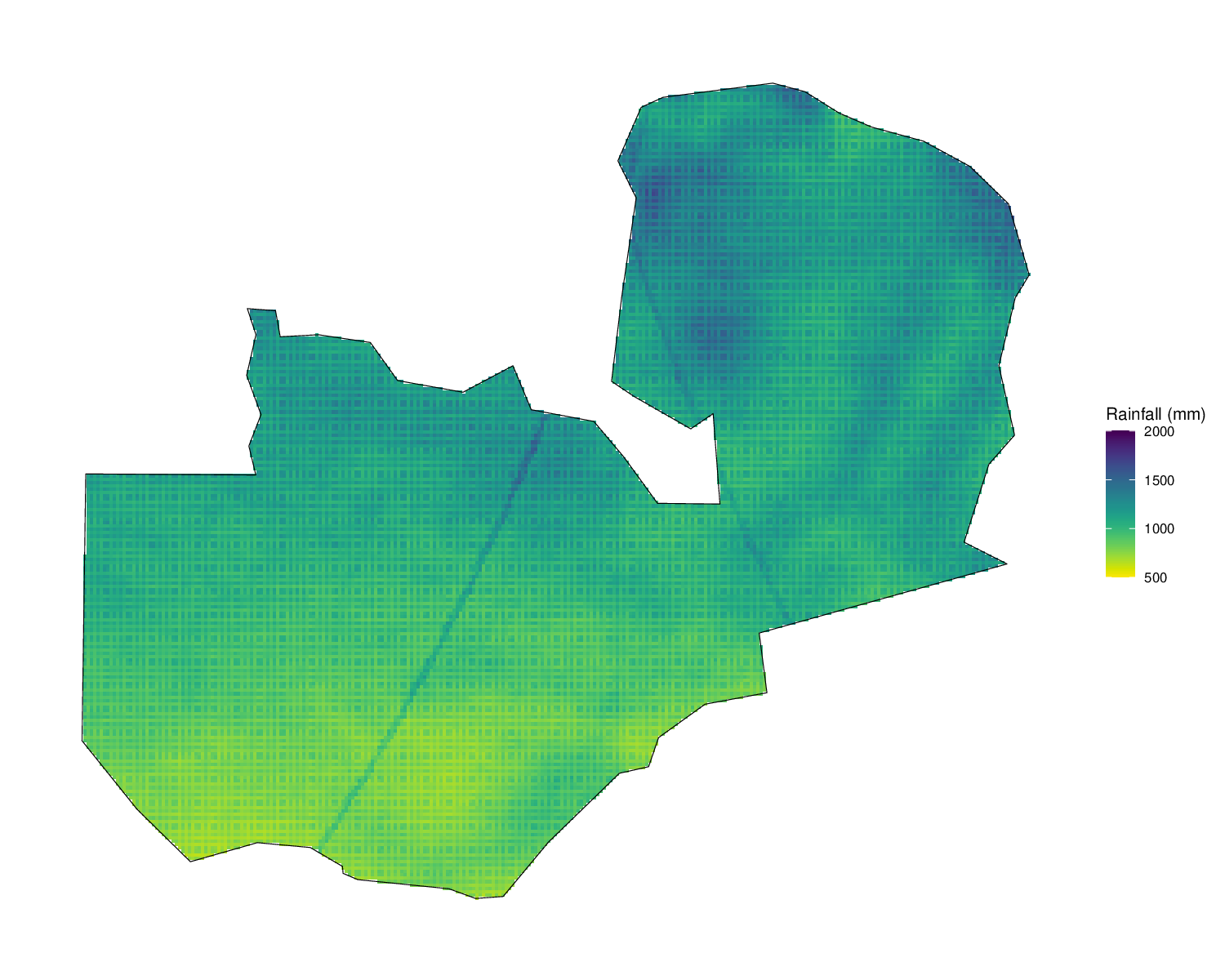}
	\caption{Mean total rainfall (mm/year) of PCCSCDR for 1983-2022 in Zambia for checking the RE's spatial consistency}\label{fig5}
\end{figure}

On the contrary, the mean annual total rainfall from the PCCSCDR product exhibits spatial irregularities characterized by localized pixelated variations, with sharp contrasts in colouration at the boundaries of the squares compared to the more uniform colouration within them, as shown in Figure \ref{fig5}. These contrasts suggest that extracting information from the dataset may yield significantly different values depending on whether the data is sourced from the boundaries or within the squares, despite their close proximity. This spatial inconsistency implies that variations in the data may arise not from actual geographical differences but rather from the pixelation effect. Consequently, such inconsistencies could lead to misleading interpretations in applications that rely on accurate spatial rainfall distribution.

\begin{figure}[ht]%
	\centering
	\includegraphics[width=0.5\textwidth]{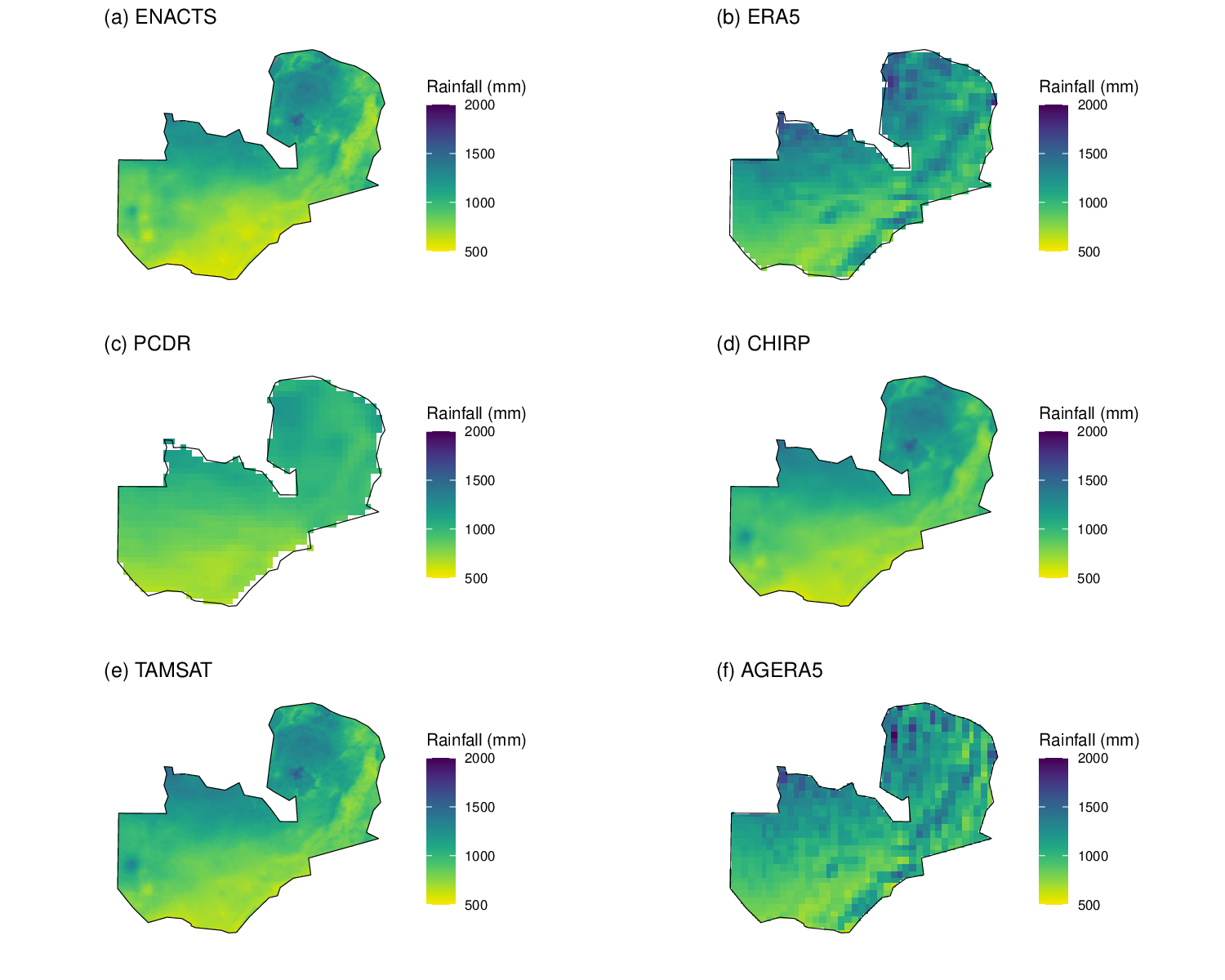}
	\caption{Mean total rainfall (mm/year) of the REs for 1983-2022 in Zambia for checking the spatial consistency of the REs}\label{fig_6}
\end{figure}

In response to the spatial inconsistency observed from PCCSCDR, we contacted the team of researchers at the Center for Hydrometeorology and Remote Sensing (CHRS) at the University of California, Irvine (UCI). They acknowledged recent observations of potential issues with the product and were investigating the underlying causes of these inconsistencies. Consequently, PCCSCDR was excluded from further consideration in this study. There are no noticeable irregularities from the other REs in Zambia, and can be seen from Figure \ref{fig_6}. 

\begin{figure}[ht]%
	\centering
	\includegraphics[width=0.5\textwidth]{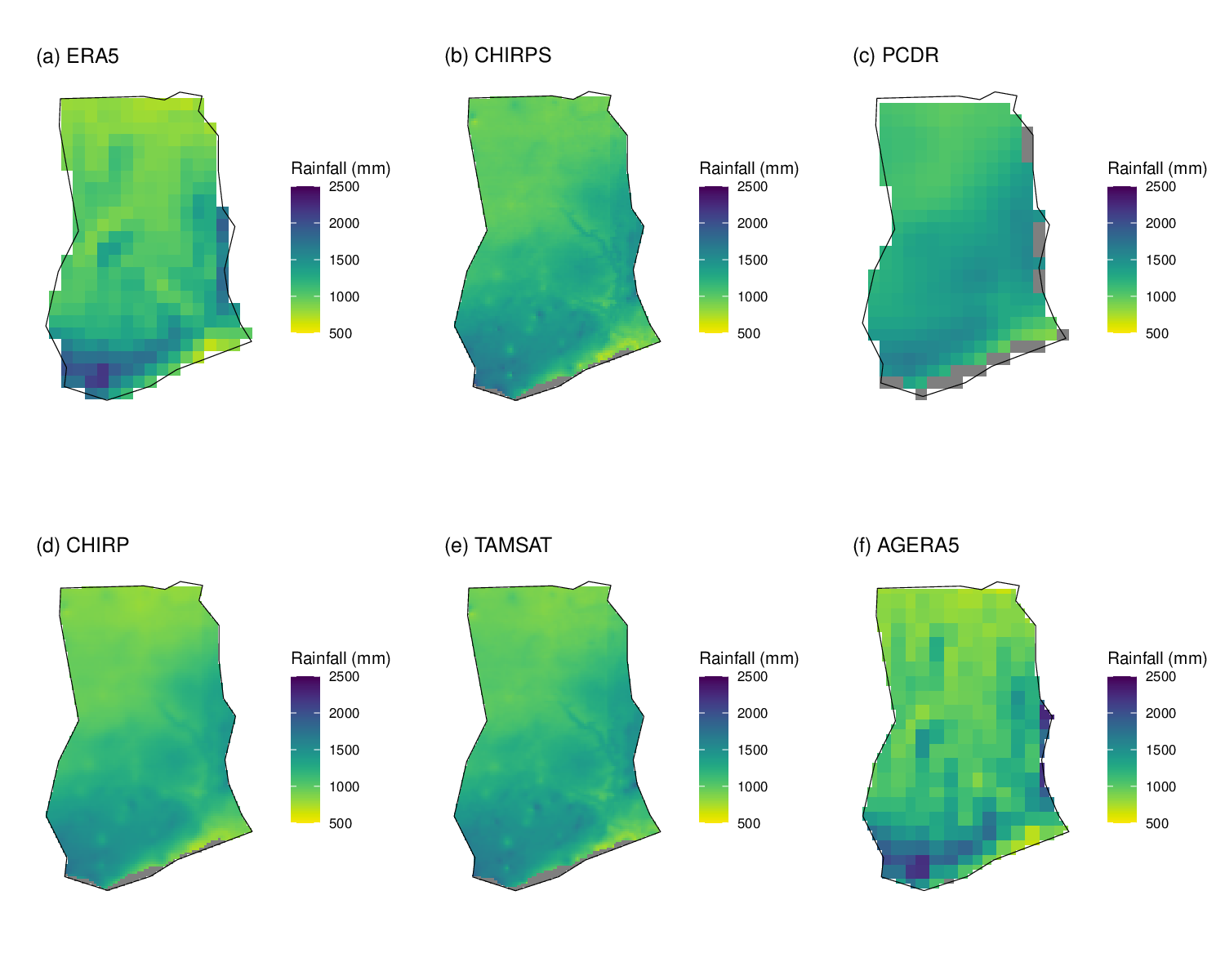}
	\caption{Mean total rainfall (mm/year) of the REs for 1983-2022 in Ghana for checking the spatial consistency of the REs}\label{fig_7}
\end{figure}

In the case of Ghana, none of the REs displayed spatial inconsistencies and were found to align closely with the country's climatology. Figure \textbf{\ref{fig_7}c} illustrates the mean annual total rainfall for PCDR, which exhibits a pixelated appearance resulting from its coarse resolution of 0.25$^{\circ}$ by 0.25$^{\circ}$ (approximately 25 km by 25 km). This pixelation is similar to that observed in ERA5 (and AGERA5 derived from ERA5), as shown in Figure \ref{fig_7}a, while CHIRPS and CHIRP (Figures \ref{fig_7}b and \ref{fig_7}d, respectively) feature a finer resolution of 0.05$^{\circ}$ by 0.05$^{\circ}$ (approximately 5 km by 5 km). Despite the pixelation, all REs effectively capture the climatological patterns in Ghana, with the northern regions and the east coast receiving less rainfall, and the central belt and west coast (particularly the Axim area) experiencing significantly higher precipitation. Overall, ENACTS, CHIRPS, CHIRP, and TAMSAT demonstrated a superior spatial consistency with the climatology of Zambia and Ghana.

\subsection{Comparison of Annual Summaries} 
The performance of the REs on annual summaries (number of rainy days, total rainfall, and mean rain per rainy day) are presented below:
\subsubsection{Number of rainy days/rain day frequency} 

\begin{figure}[ht]%
	\centering
	\includegraphics[width=0.5\textwidth]{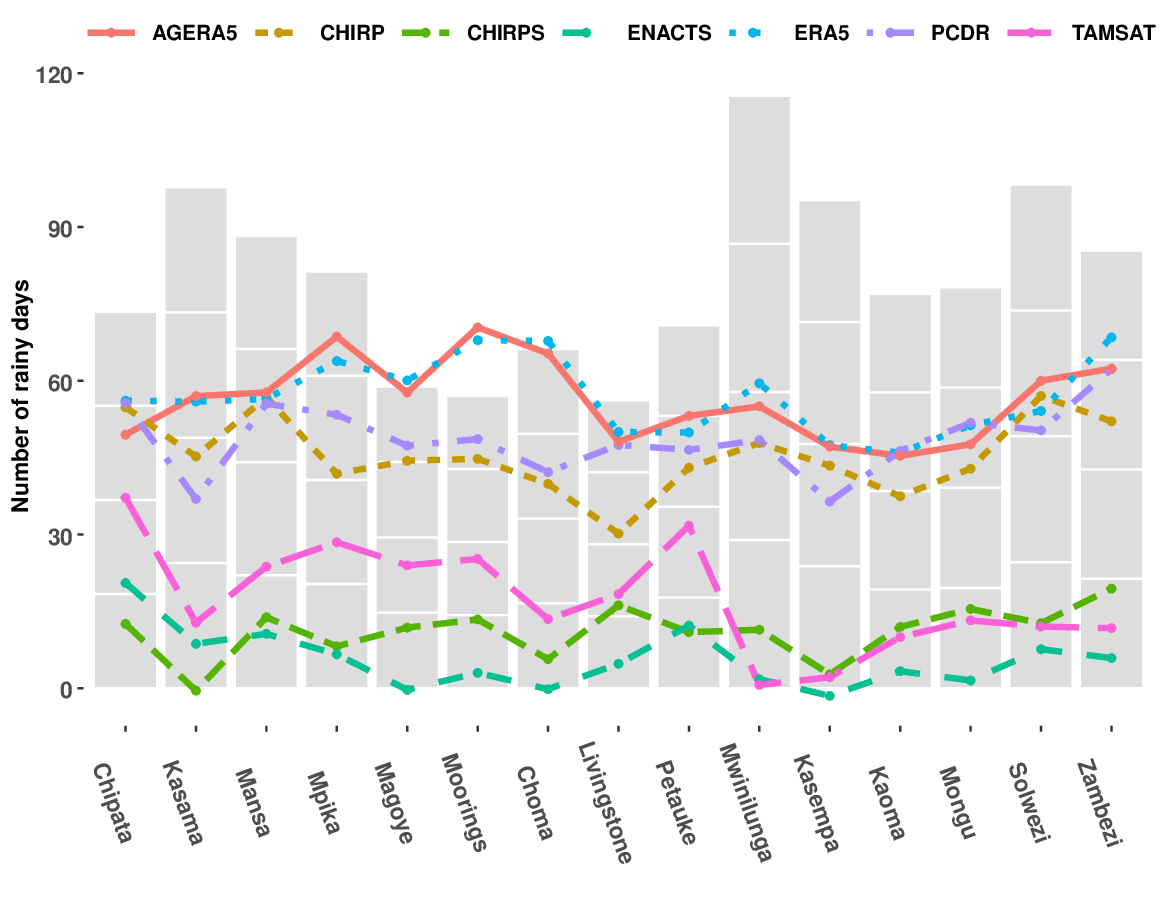}
	\caption{ME of number of rainy days in Zambia for each RE on the quartiles of the number of rainy days at each station}\label{fig_8}
\end{figure}
\begin{figure}[ht]%
	\centering
	\includegraphics[width=0.5\textwidth]{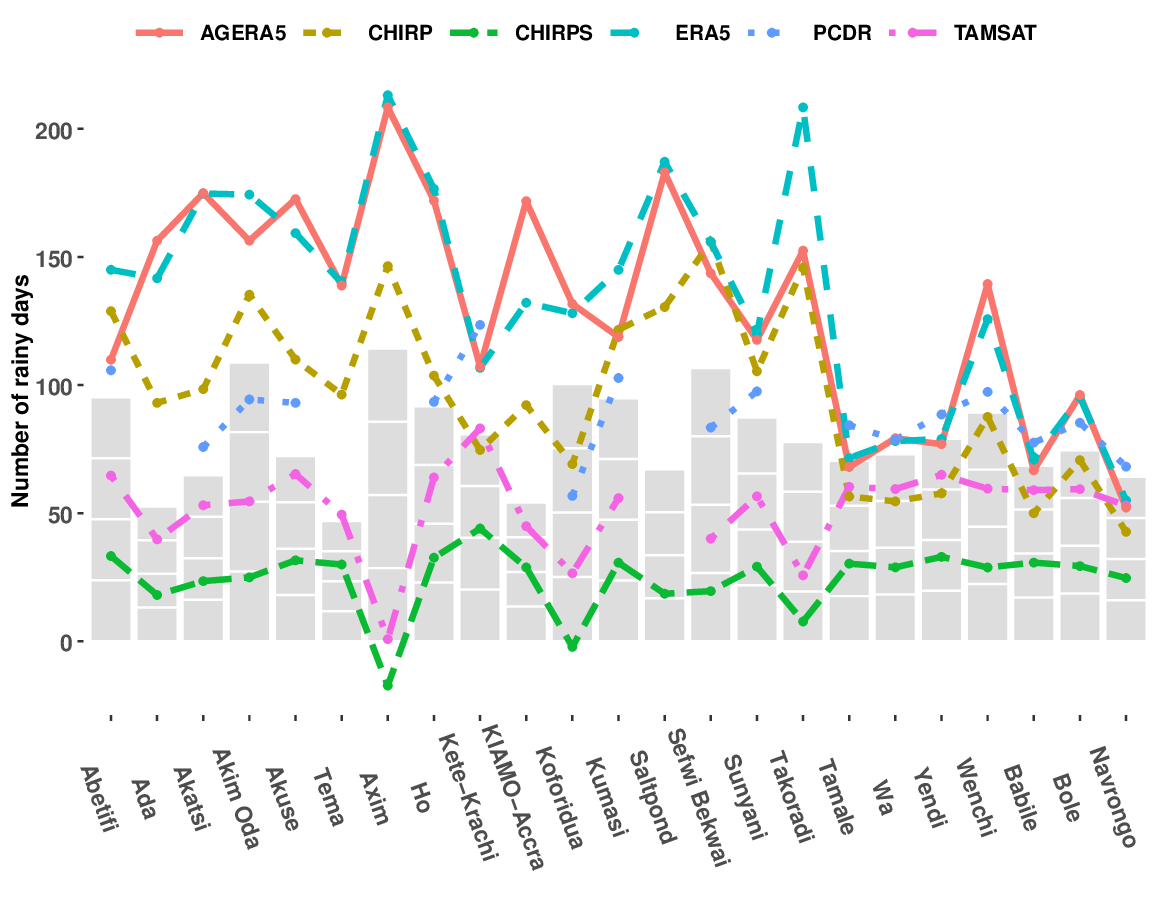}
	\caption{ME of number of rainy days in Ghana for each RE on the quartiles of the number of rainy days at each station}\label{fig_9}
\end{figure}

The line plots in Figures \ref{fig_8} and \ref{fig_9}show the MEs on the number of rainy days of each product at the various stations in Zambia and Ghana respectively. The height of the bars at the different stations represents the mean number of rainy days at those stations. Each bar is divided into four equal segments (separated by horizontal white lines), with the first segment at the bottom and the fourth segment at the top. These divisions facilitate the estimation of PBIAS. For example, TAMSAT has an ME of about 40 mm at Chipata while having a PBIAS of about 50\% (see Figure \ref{fig_8}).

On average, all the REs overestimated the number of rainy days in Zambia and Ghana (Figures~\ref{fig_8} and \ref{fig_9}).

Among these products, ENACTS showed the lowest average bias in Zambia, followed by CHIRPS and TAMSAT. Apart from ENACTS, CHIRPS, and TAMSAT, the remaining products exhibited an average bias exceeding 50\% in Zambia, indicating a considerable overestimation of rainy days. 

The overestimation was higher in Ghana. With the exception of CHIRPS and TAMSAT, the remaining REs exhibited an average bias exceeding 100\% (Figure \ref{fig_9}). For CHIRPS, TAMSAT, and, to a lesser extent, PCDR, the overestimation tended to be more pronounced in the northern stations (Tamale, Wa, Yendi, Babile, Bole, and Navrongo) than at the southern stations. On the contrary, the overestimation by the other REs was more pronounced, on average, at the southern stations than the northern stations. 

\begin{figure}[ht]%
	\centering \includegraphics[width=0.5\textwidth]{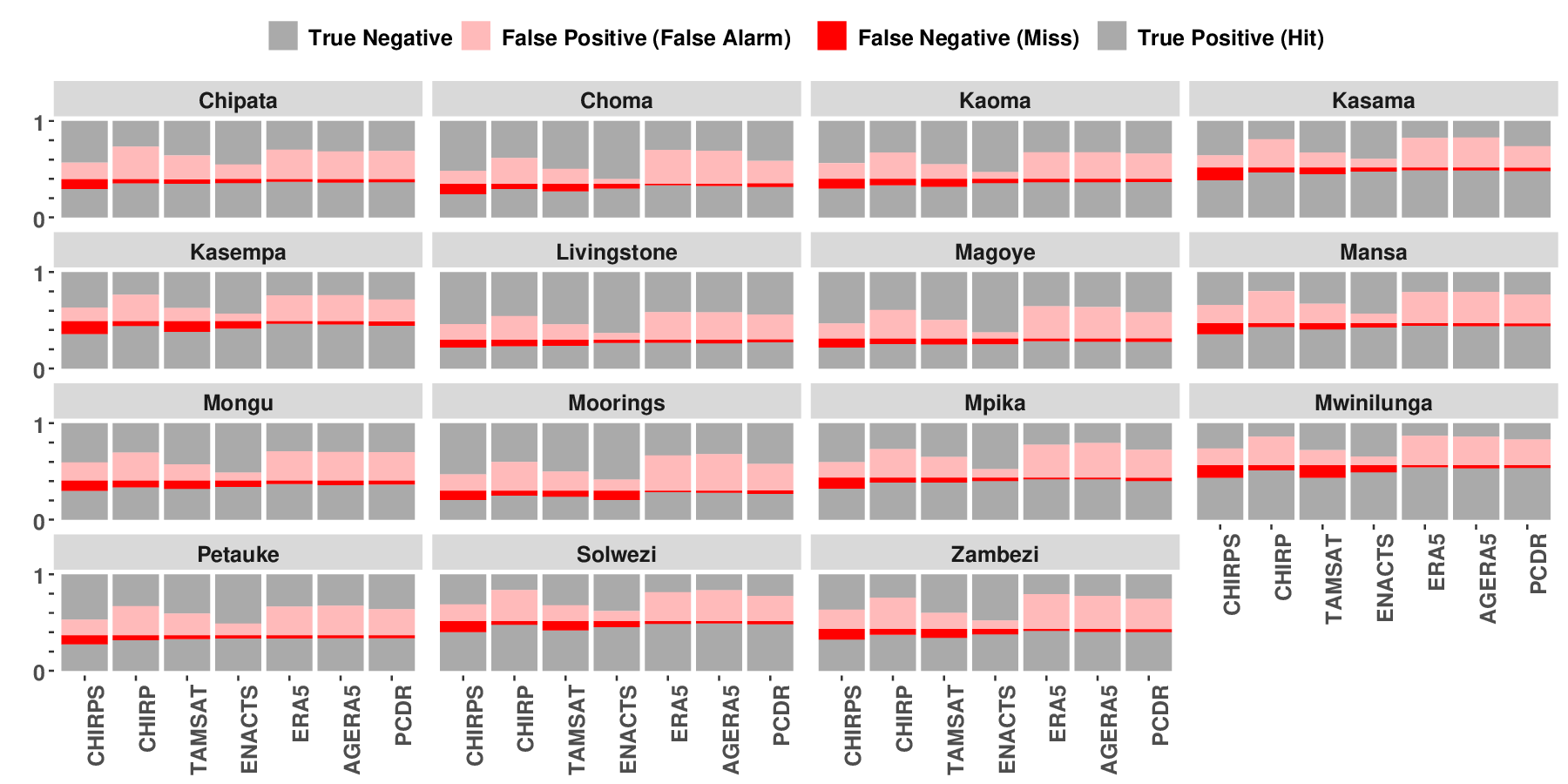} \caption{Stacked bar chart illustrating the performance of different REs in detecting rainy days across various stations in Zambia, as represented in the graph facets.}\label{fig_10} 
\end{figure}

\begin{figure}[ht]%
	\centering
	\includegraphics[width=0.5\textwidth]{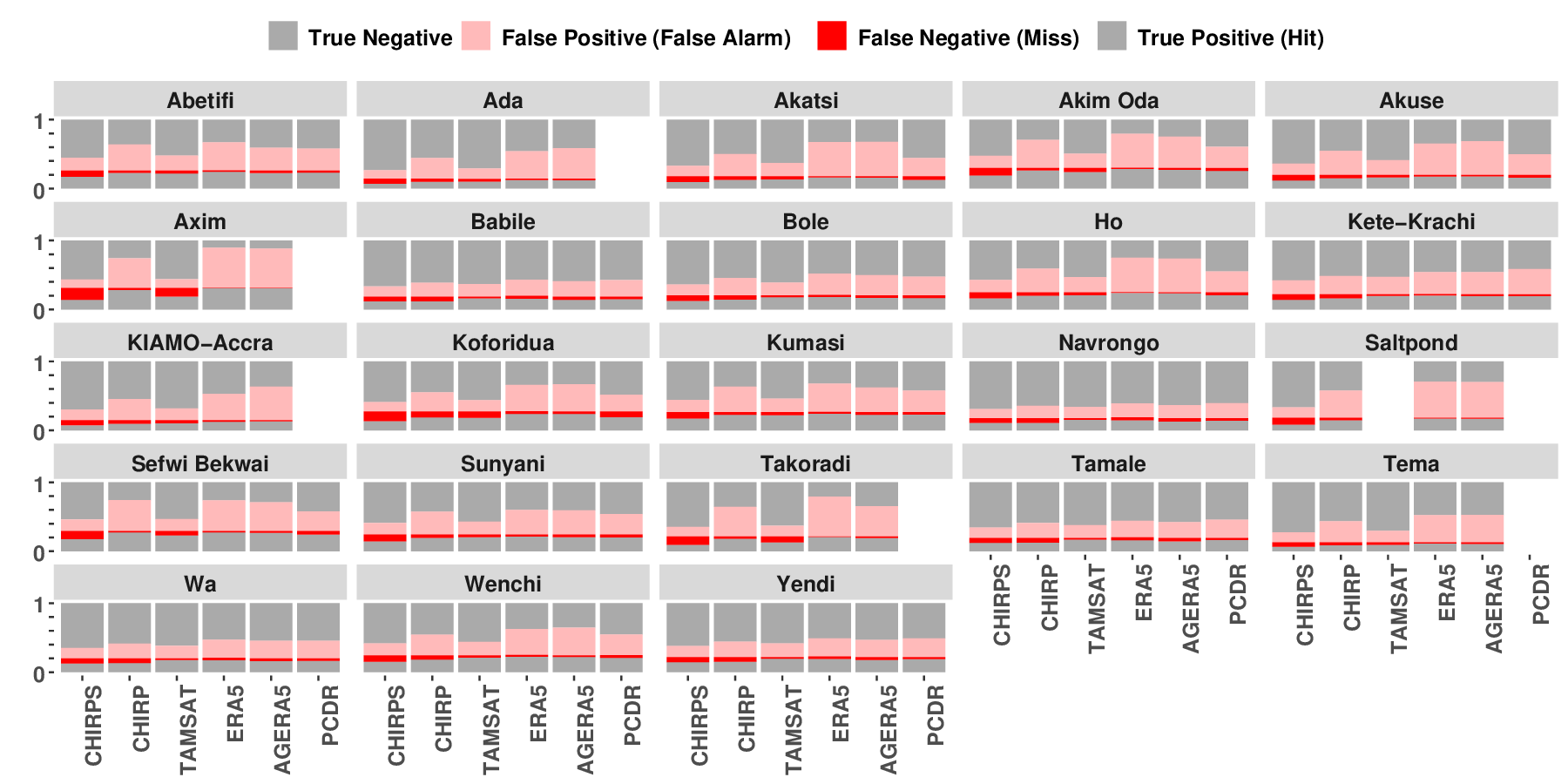}
	\caption{A stacked bar chart showing how well the REs detect rainy days at the different stations in Ghana (shown in different facets of the graph)}\label{fig_11}
\end{figure}

Figures~\ref{fig_10} and \ref{fig_11} illustrate the performs of the REs in detecting rainy days across various stations in Zambia and Ghana respectively.

From Figures~\ref{fig_10} and \ref{fig_11}, it was observed that all the products had considerable proportions of False Positives. This occurrence was lowest in ENACTS, followed by CHIRPS and TAMSAT. Proportion of False Negatives were lowest in ERA5, AGERA5, and PCDR and was relatively higher in CHIRPS, ENACTS, and CHIRP. 

\subsubsection{Annual rainfall totals}
\begin{figure}[ht]%
	\centering
	\includegraphics[width=0.5\textwidth]{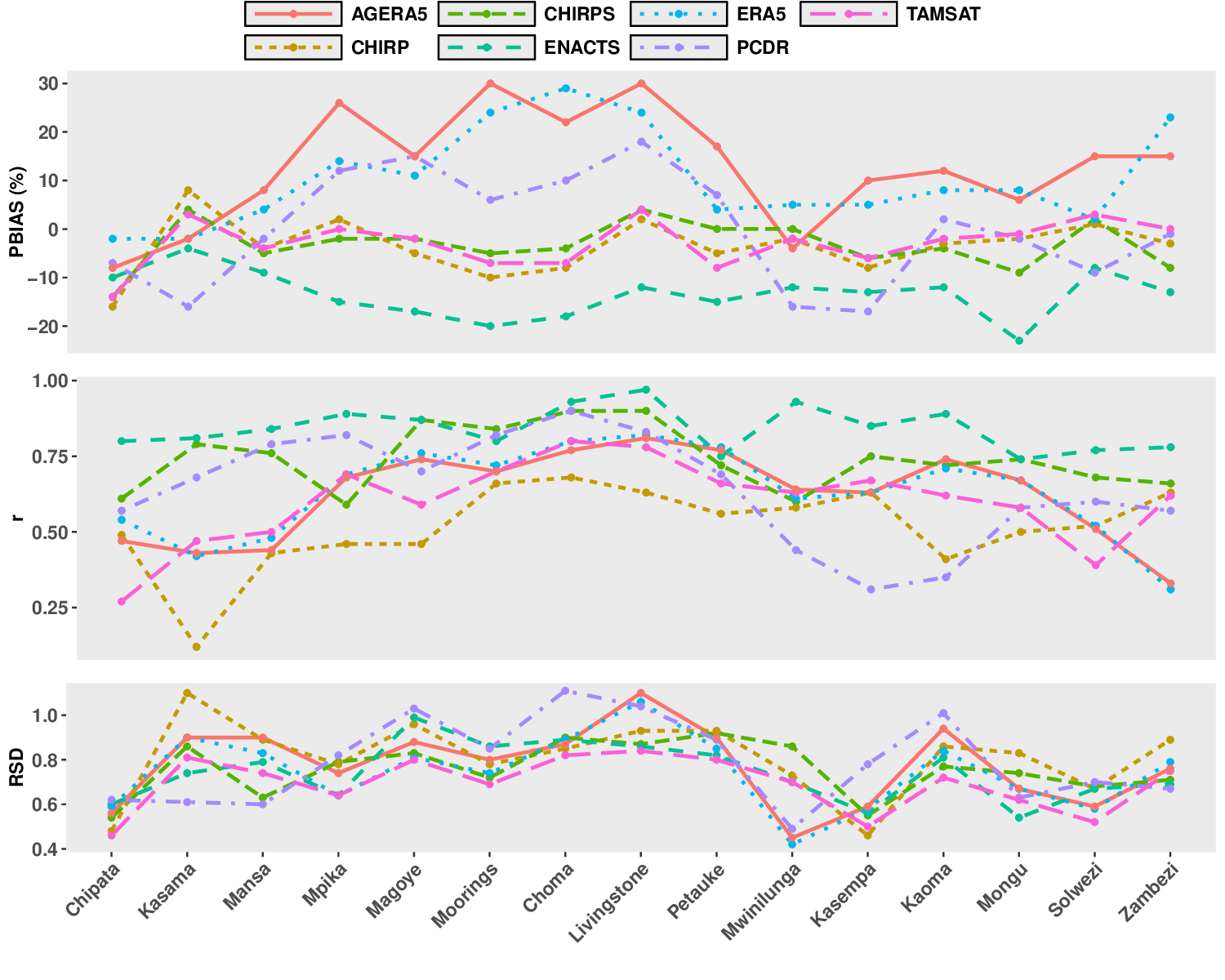}
	\caption{Line plots showing PBIAS, r, and RSD on annual total rainfall for all the products (in different colours and line types) across all the stations (on the x-axis) in Zambia. The y-axis displays the values of the corresponding metric}\label{fig_12}
\end{figure}

\begin{figure}[ht]%
	\centering
	\includegraphics[width=0.5\textwidth]{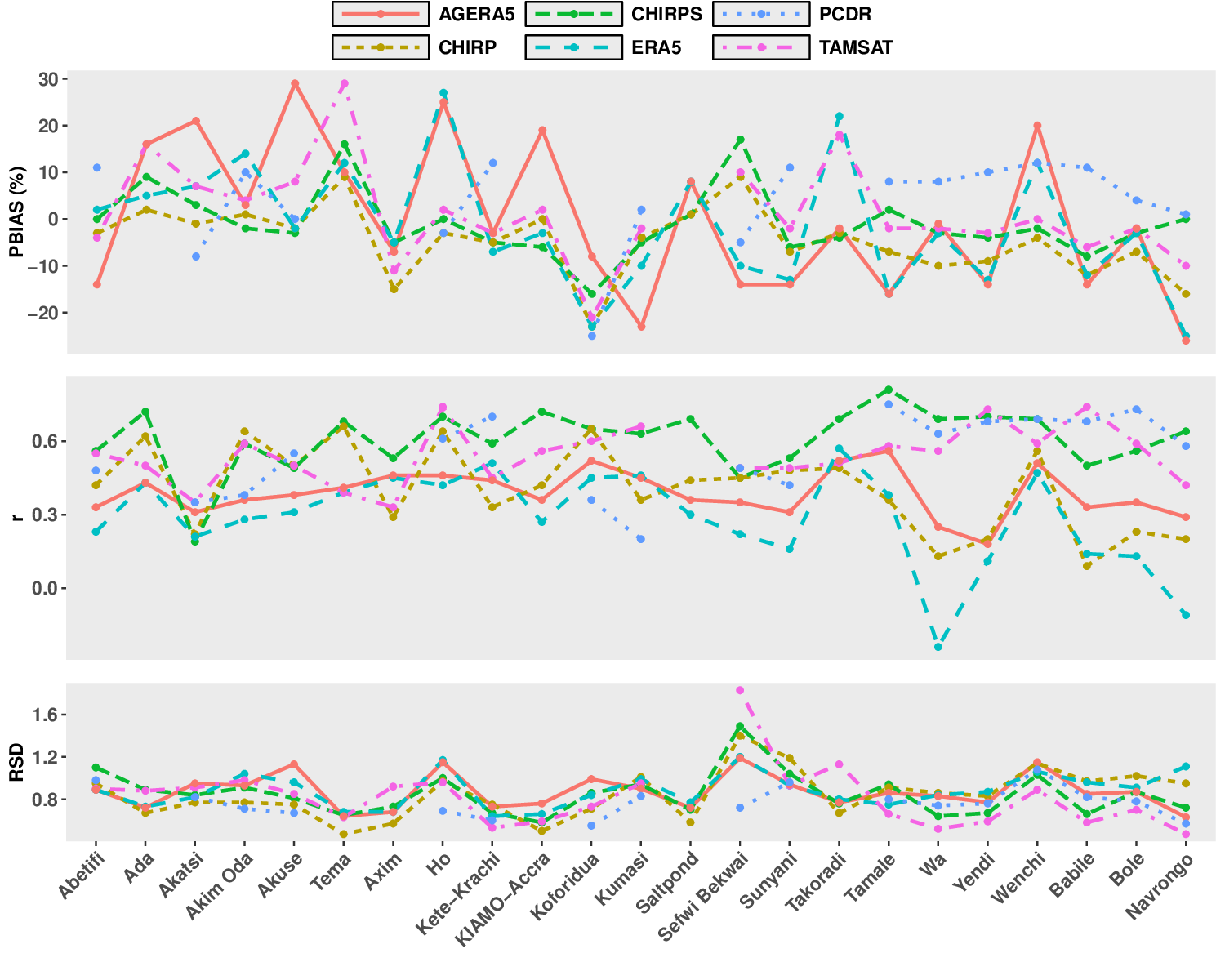}
	\caption{Line plots showing PBIAS, r, and RSD on annual total rainfall for all the products (in different colours and line types) across all the stations (on the x-axis) in Ghana. The y-axis displays the values of the corresponding metric}\label{fig_13}
\end{figure}

As compared to rainfall frequency, the biases in total rainfall were generally low for all the REs both in Zambia and Ghana ($-30$\% $\le$ PBIAS $\le$ 30\%) as can be seen in Figures \ref{fig_12} and \ref{fig_13} respectively. CHIRPS, TAMSAT and CHIRP had the least biases ($-20$\% $<$ PBIAS $\le$ 10\%) in Zambia. ENACTS, CHIRPS, CHIRP, and TAMSAT underestimated total annual rainfall at most of the stations in Zambia while ERA5 and AGERA5, overestimated total rainfall at most of the stations in Zambia (Figure~\ref{fig_12}). PCDR overestimated at some of the stations (0\% $<$ PBIAS $<$ 20\%) while underestimating at some (with $-20$\% $<$ PBIAS $<$ 0\%). The product with the highest underestimation was ENACTS ($-30$\% $\le$ PBIAS $<$ 0\%) across all stations while AGERA5 was the product with the highest overestimation (0\% $<$ PBIAS $\le$ 30\%) at most of the stations. 

The total rainfall of the REs tended to correlate well with observed data (0.5 $<$ r $<$ 1) across most of the stations in Zambia (Figure \ref{fig_12}). ENACTS seemed to have the highest correlations across the stations, followed by CHIRPS. CHIRP appeared to have the lowest correlations, generally less than 0.6. While CHIRPS and TAMSAT appeared to correlate well with gauge observations in Ghana (Figure \ref{fig_13}), with 0.5 $<$ r $<$ 1, the correlation from the other products were not as good (r $<$ 0.6 at most of the stations). ERA5 had the lowest correlations, followed by AGERA5 and CHIRP.  

The REs tended to have less variability than the observed rainfall total (RSD $<$ 1) across most of the stations in Zambia and Ghana (Figures~\ref{fig_12} and \ref{fig_13} respectively).

\subsubsection{Mean rain per rainy day}

\begin{figure}[ht]%
	\centering
	\includegraphics[width=0.5\textwidth]{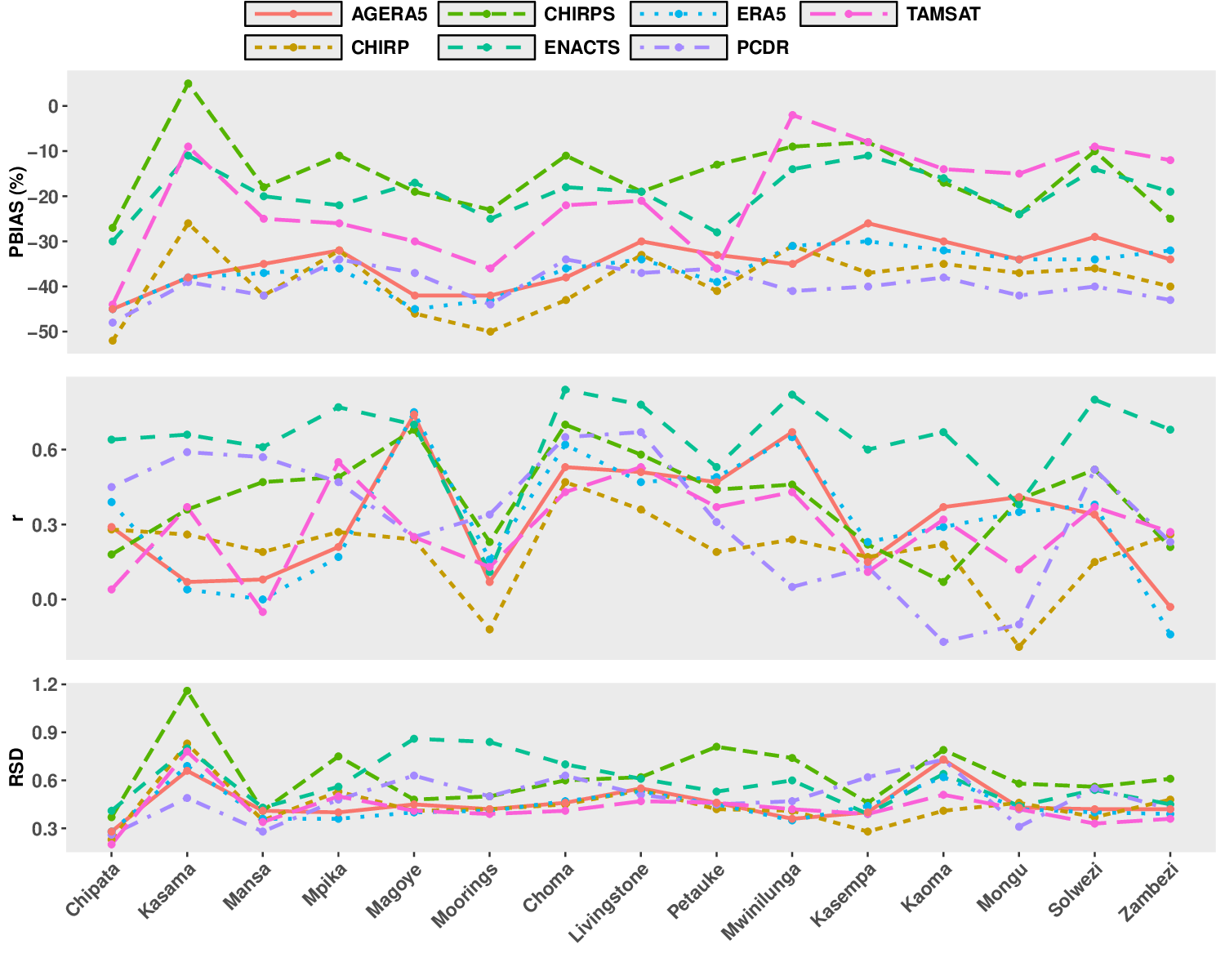}
	\caption{Line plots showing PBIAS, r, and RSD on mean rain per rainy day for all the products (in different colours and line types) across all the stations (on the x-axis) in Zambia. The y-axis displays the values of the corresponding metric}\label{fig_14}
\end{figure}

\begin{figure}[ht]%
	\centering
	\includegraphics[width=0.5\textwidth]{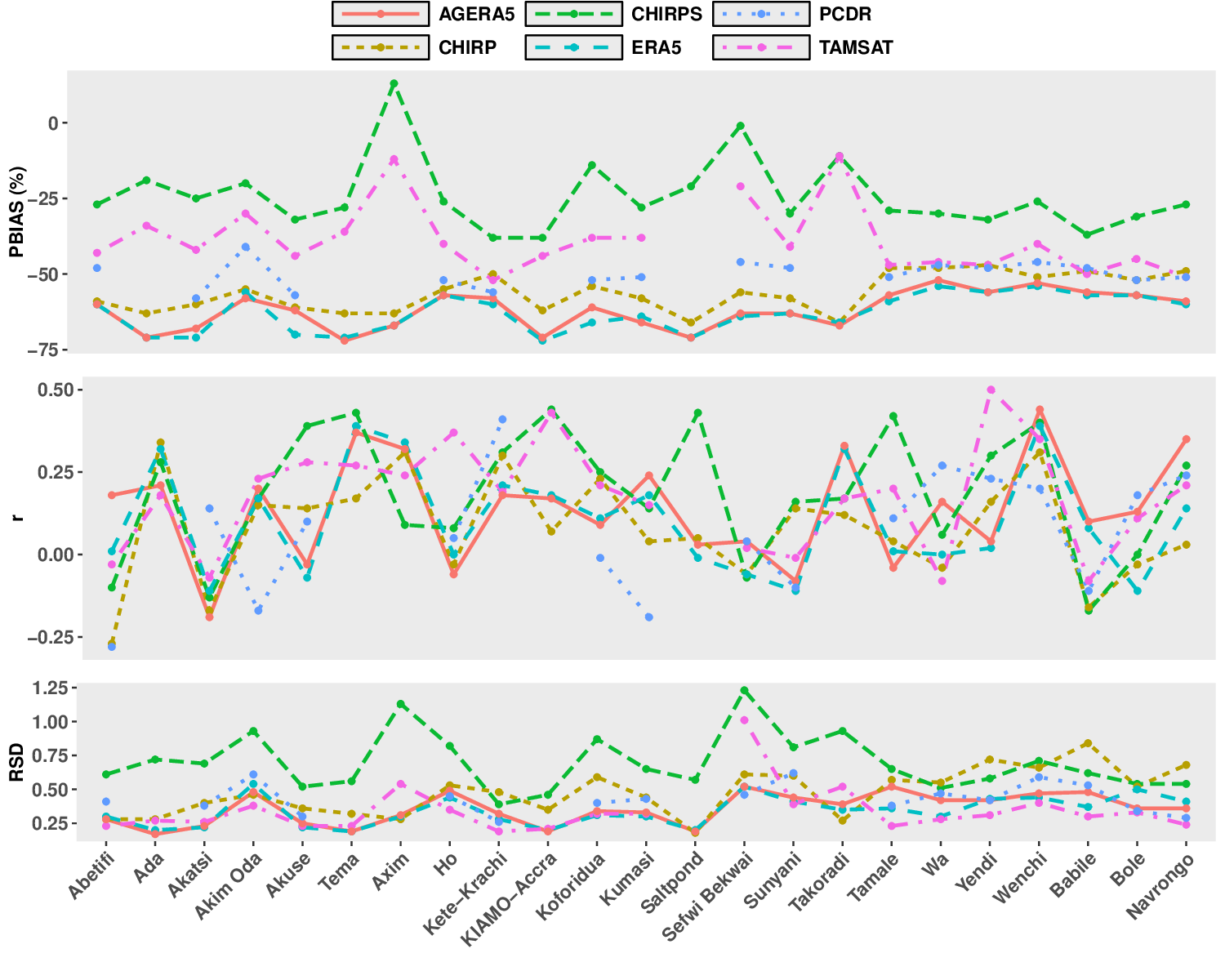}
	\caption{Line plots showing PBIAS, r, and RSD on mean rain per rainy day for all the products (in different colours and line types) across all the stations (on the x-axis) in Ghana. The y-axis displays the values of the corresponding metric}\label{fig_15}
\end{figure}

The REs underestimated mean rain per rainy day at varying degrees, with $-60$\% $<$ PBIAS $<$ 0\% at most of the stations in Zambia (Figure \ref{fig_14}), and $-75$\% $<$ PBIAS $<$ 0 at most of the stations in Ghana (Figure~\ref{fig_15}). Correlations were generally low for most of the products across all the stations. Aside from ENACTS (with 0.5 $<$ r $<$ 1 at most of the locations), most of the REs had correlations between $-0.3$ and 0.5 at most of the stations in Zambia and Ghana (respectively Figures~\ref{fig_14} and \ref{fig_15}). The REs also tended to have lesser variability (RSD $<$ 0.8) across most of the stations as compared to gauge observations in Zambia and Ghana.

\subsection{Comparison of Seasonal Behaviour}
\begin{figure}[ht]%
	\centering
	\includegraphics[width=0.5\textwidth]{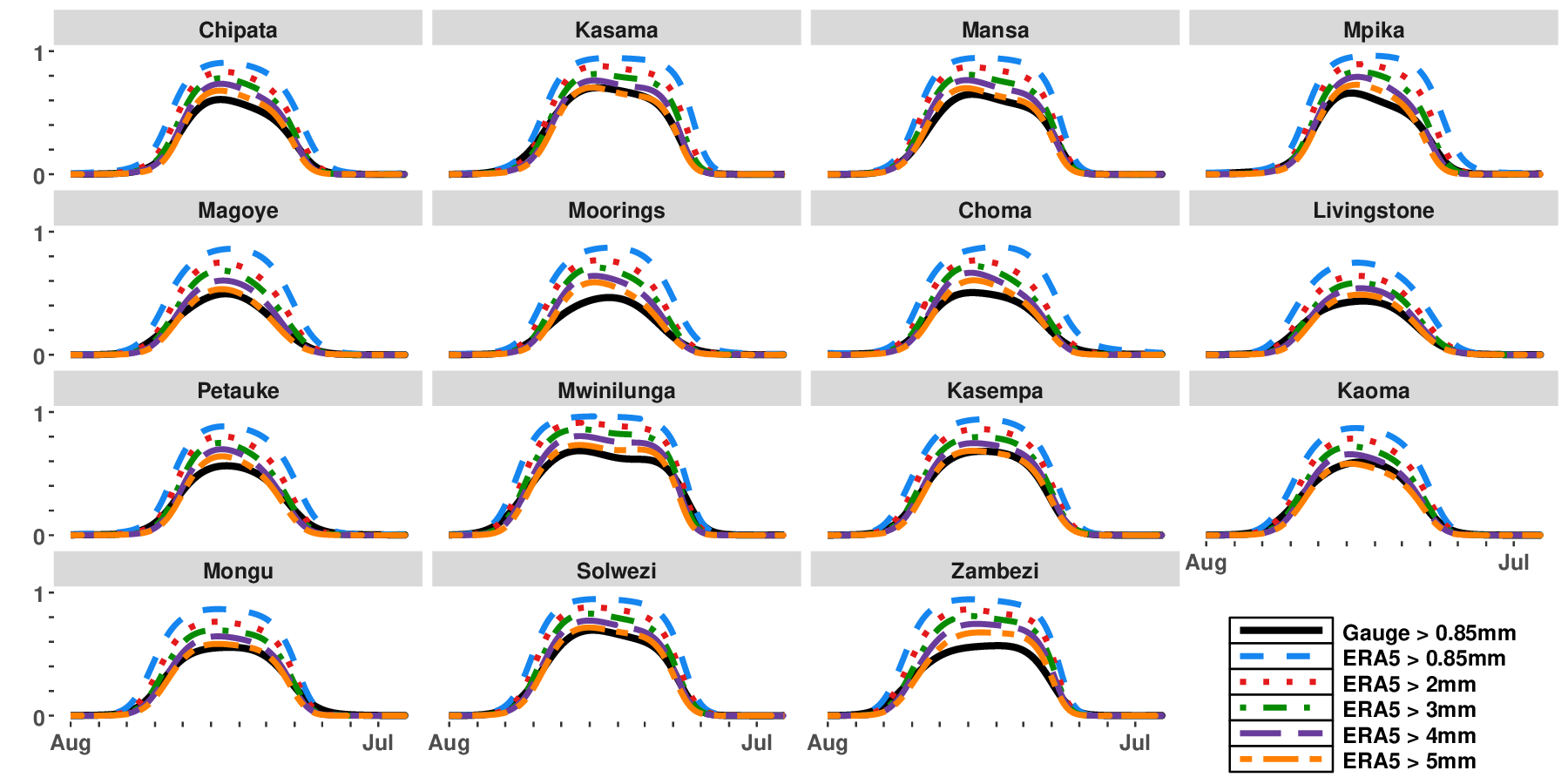}
	\caption{Rain day frequency models of gauge observations vs ERA5 in Zambia. At each station, only days which the gauge observations and RE values were not missing were included to ensure comparison of the same days. The solid black curve shows the fitted rain day frequency for the gauge observations with 0.85mm rain day threshold. The coloured dashed lines show the fitted rain day frequency for the RE values at various rain day thresholds (0.85mm, 2mm, 3mm, 4mm, and 5mm). The y-axis on each plot represents the proportion of rain day frequency, while the x-axis shows the date in the year, spanning from August 1 to July 31. Figures~\ref{fig_17}-\ref{fig_22} were created in a similar way except that a year at the Ghanaian stations start from January 1 and end on December 31}\label{fig_16}
\end{figure}

\begin{figure}[ht]%
	\centering
	\includegraphics[width=0.5\textwidth]{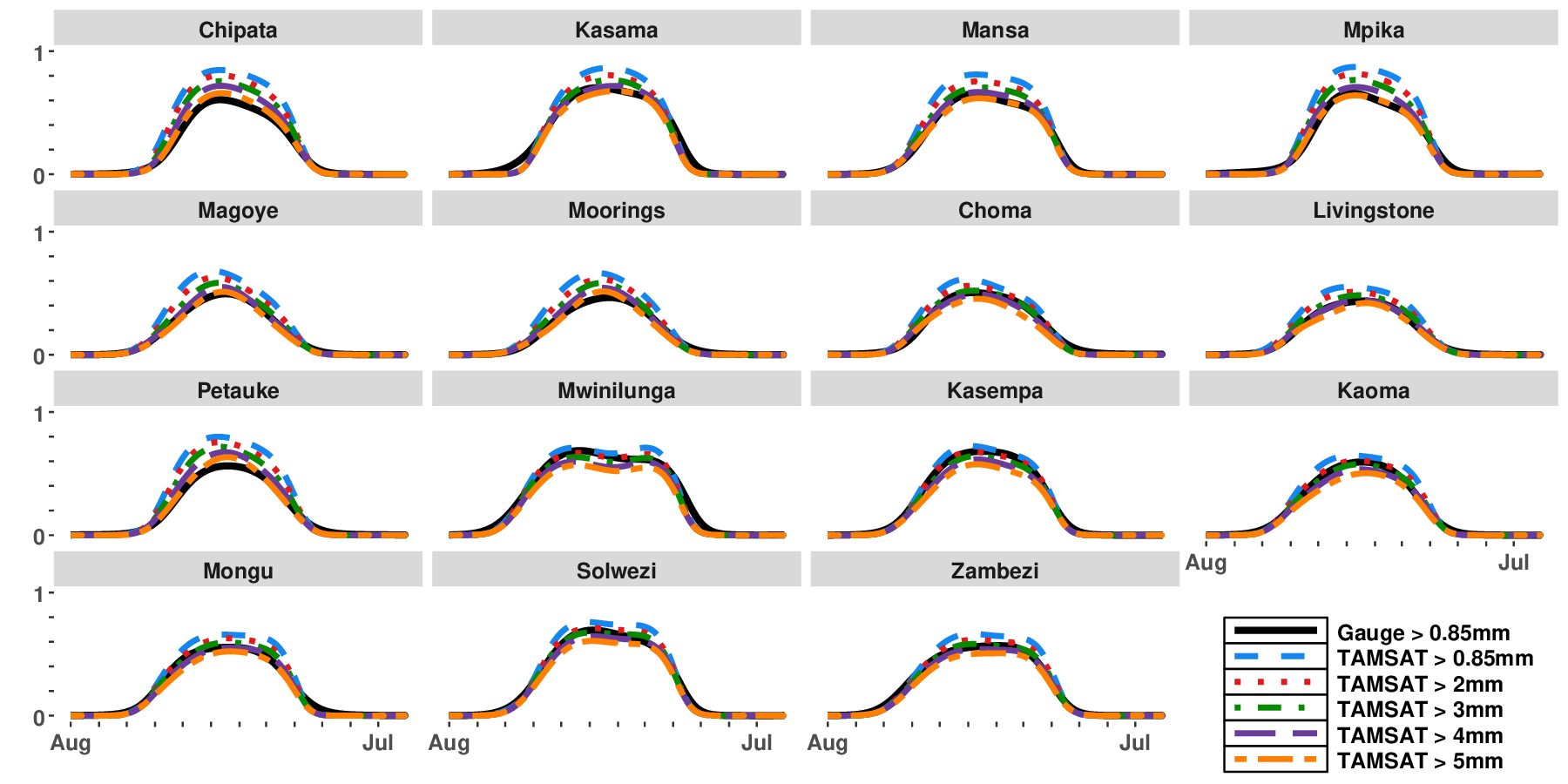}
	\caption{Rain day frequency models of gauge observations vs TAMSAT in Zambia }\label{fig_17}
\end{figure}

\begin{figure}[ht]%
	\centering
	\includegraphics[width=0.5\textwidth]{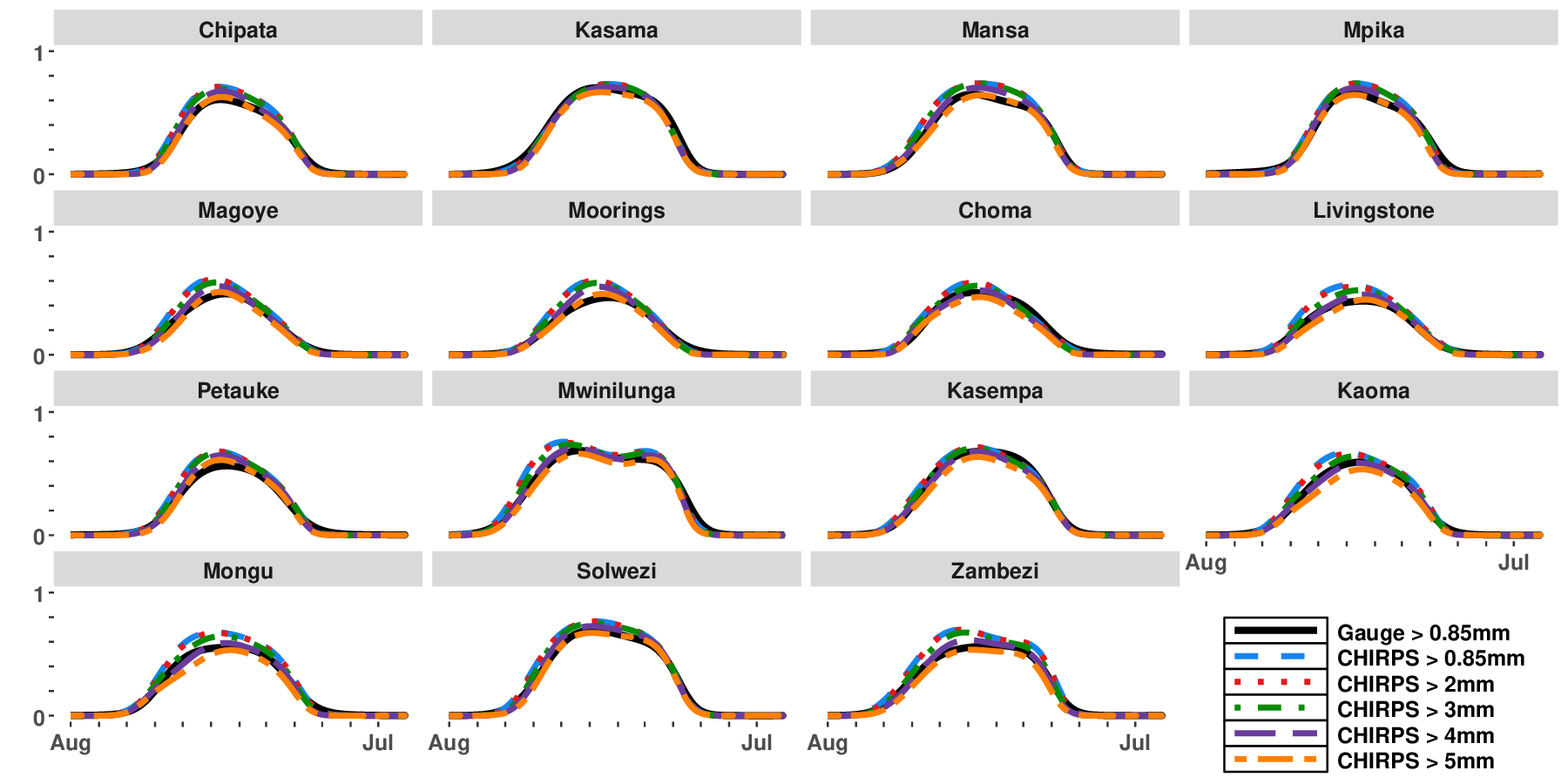}
	\caption{Rain day frequency models of gauge observations vs CHIRPS in Zambia}\label{fig_18}
\end{figure}

\begin{figure}[ht]%
	\centering
	\includegraphics[width=0.5\textwidth]{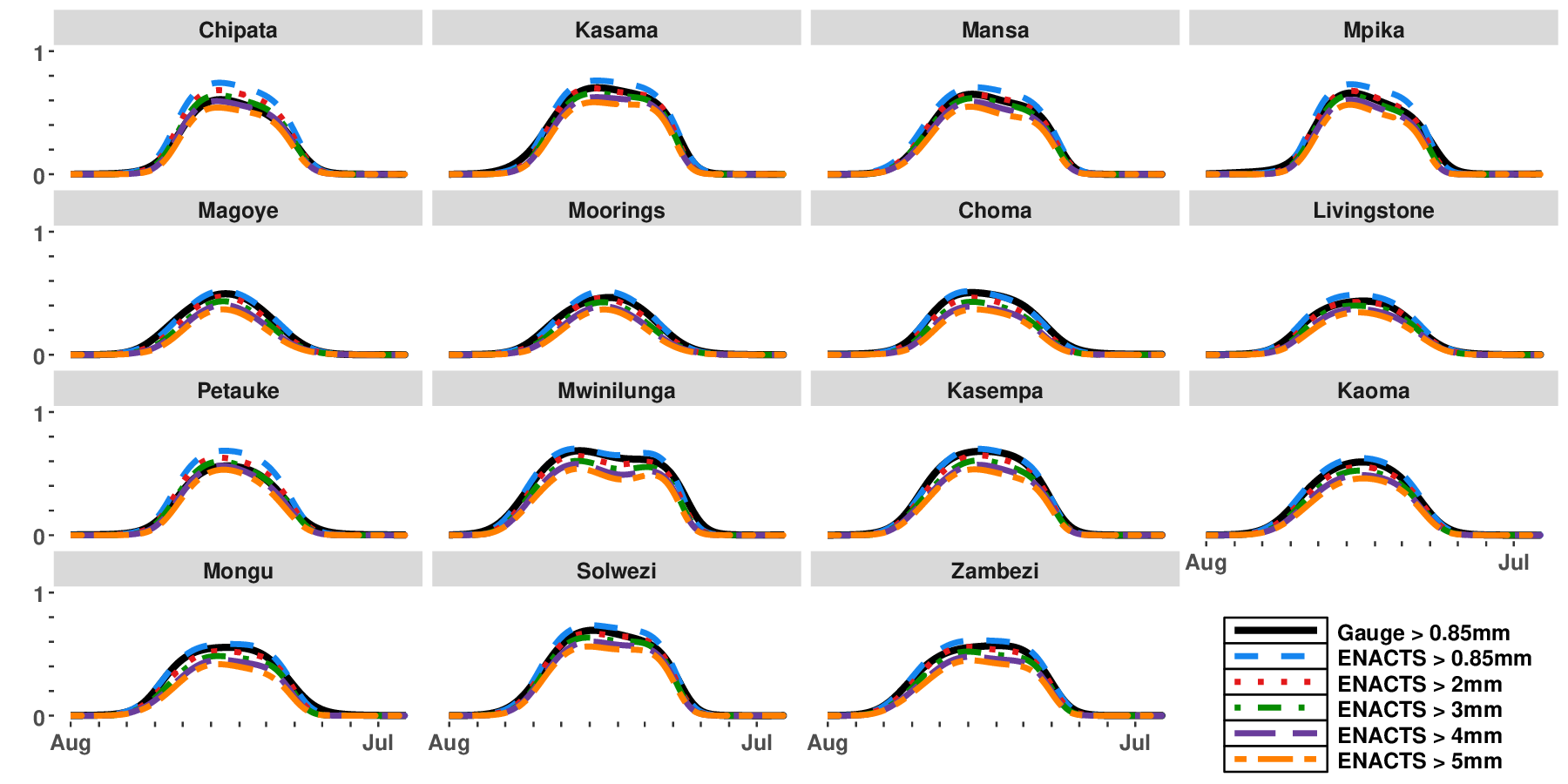}
	\caption{Rain day frequency models of gauge observations vs ENACTS in Zambia}\label{fig_19}
\end{figure}

\begin{figure}[ht]%
	\centering
	\includegraphics[width=0.5\textwidth]{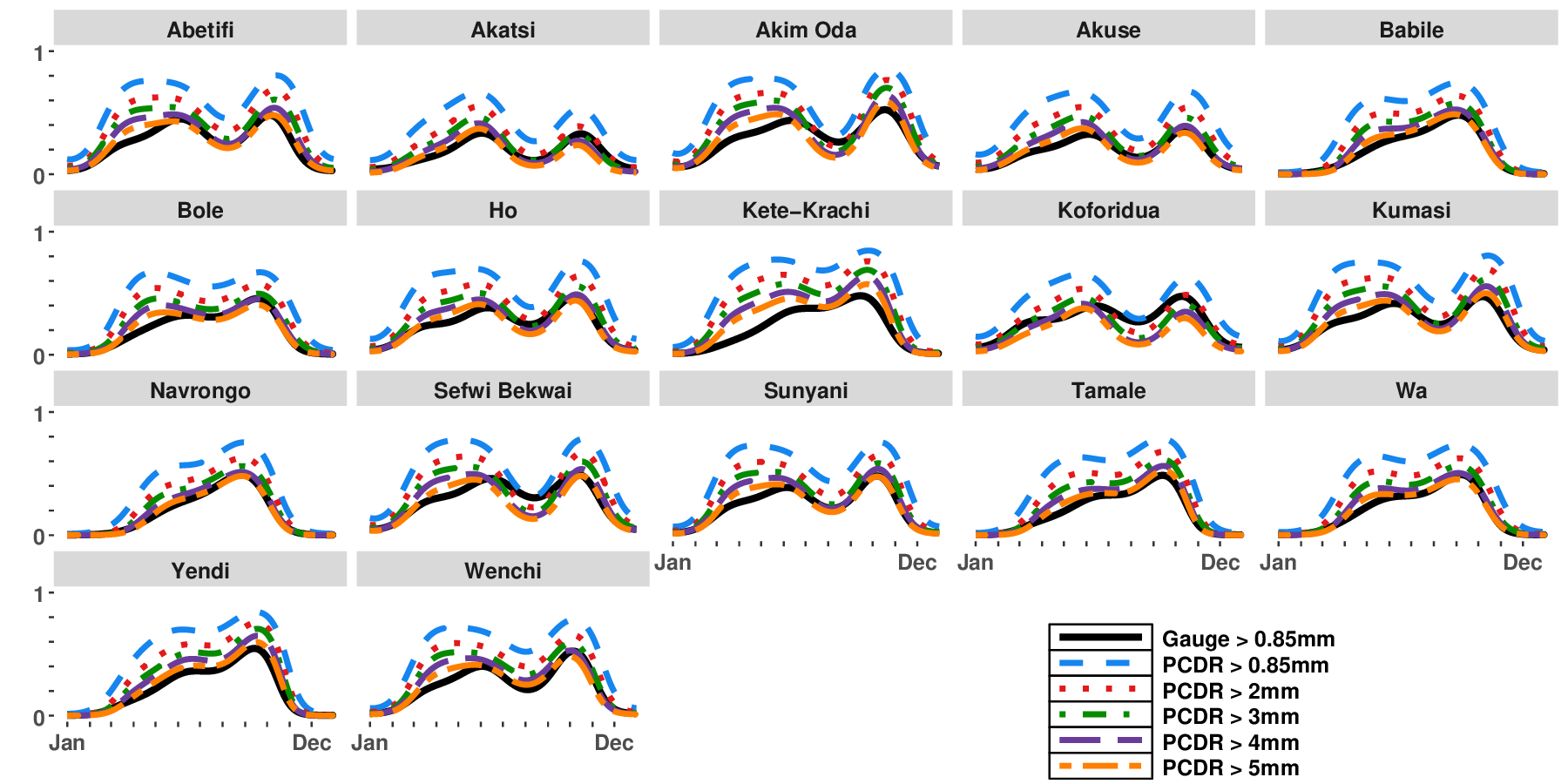}
	\caption{Rain day frequency models of gauge observations vs PCDR in Ghana}\label{fig_20}
\end{figure}

\begin{figure}[ht]%
	\centering
	\includegraphics[width=0.5\textwidth]{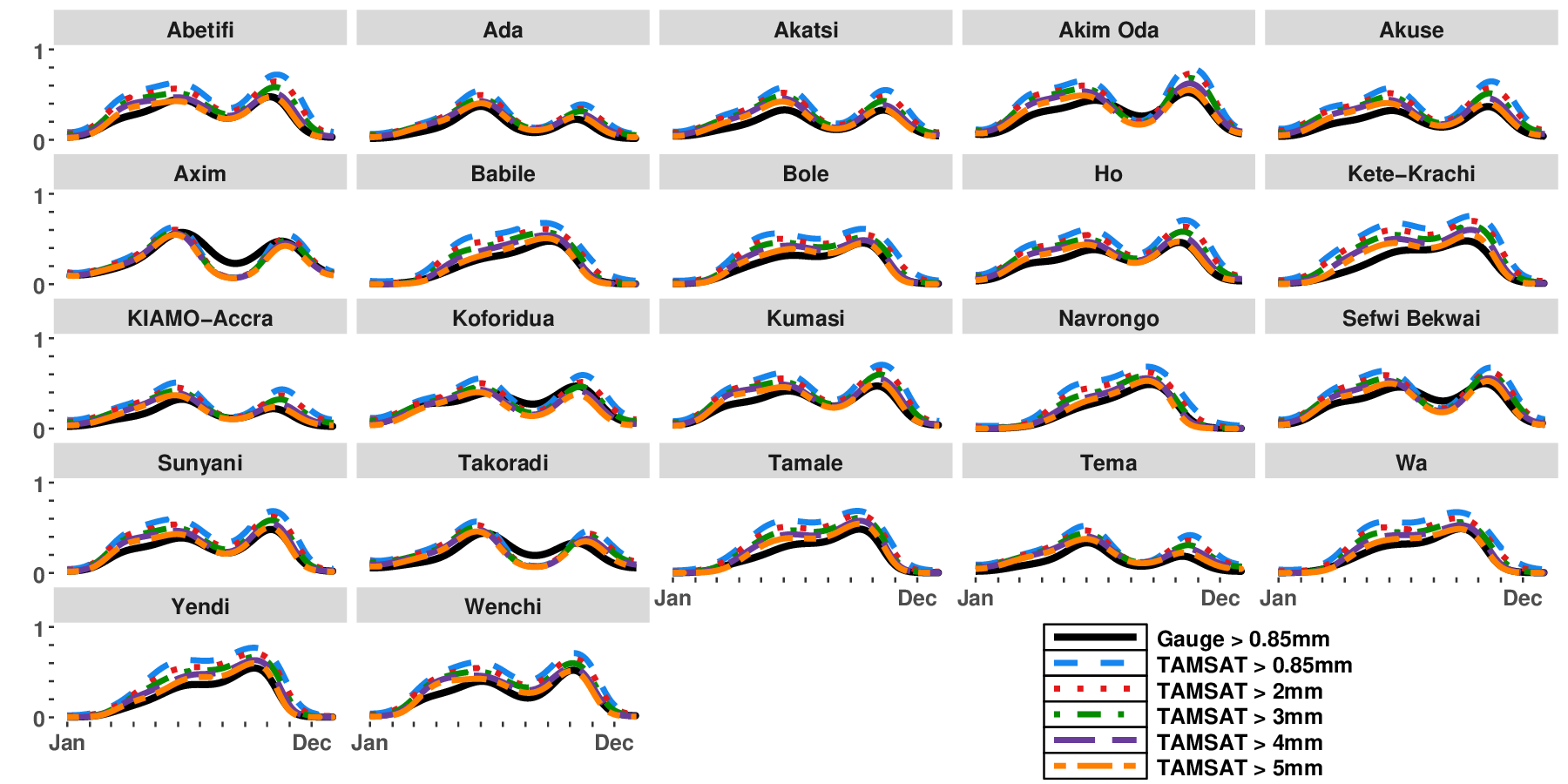}
	\caption{Rain day frequency models of gauge observations vs TAMSAT in Ghana}\label{fig_21}
\end{figure}

\begin{figure}[ht]%
	\centering
	\includegraphics[width=0.5\textwidth]{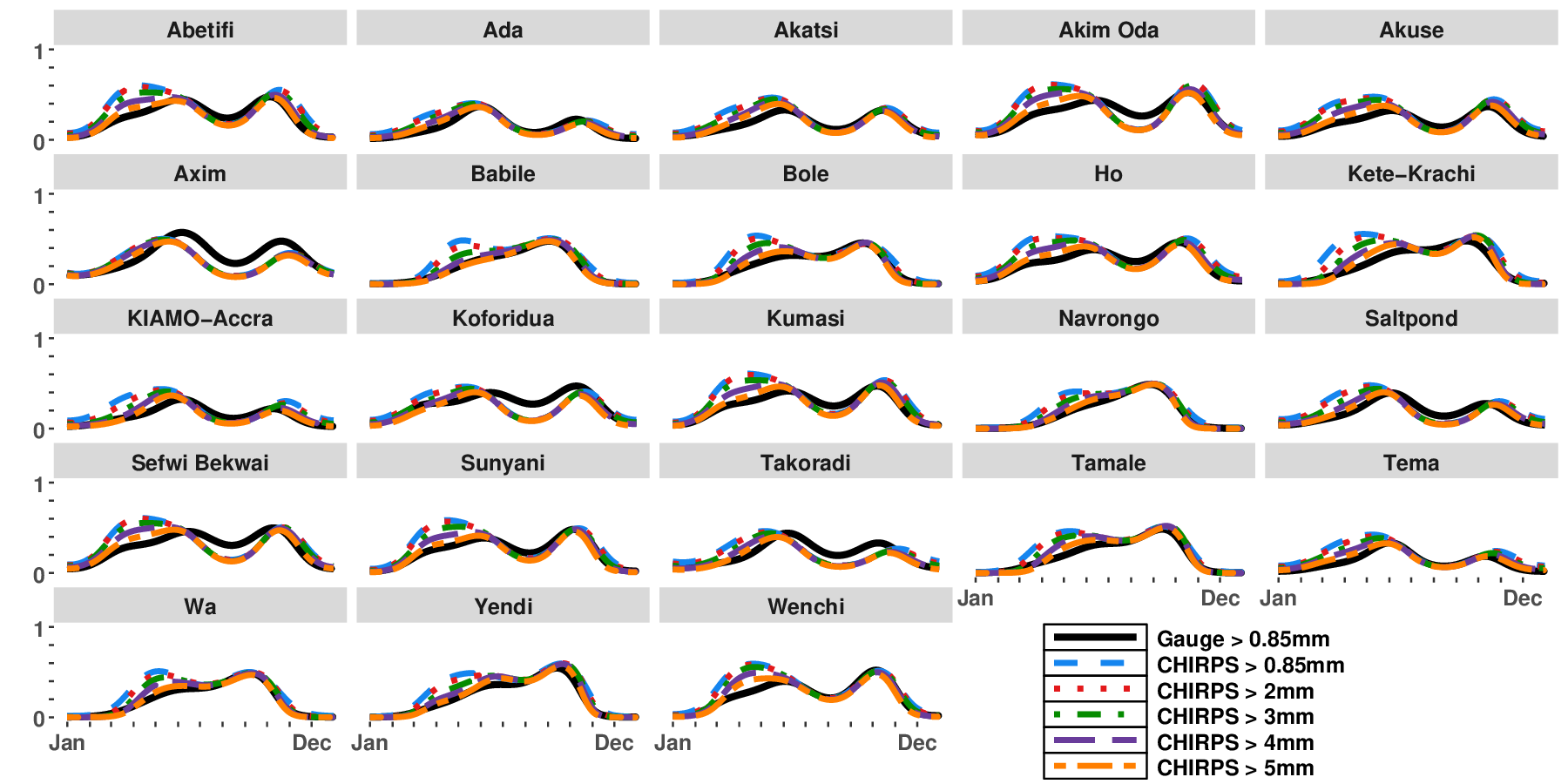}
	\caption{Rain day frequency models of gauge observations vs CHIRPS in Ghana}\label{fig_22}
\end{figure}

Figures \ref{fig_16}-\ref{fig_22} show the Zero-order Markov chain models with three harmonics fitted separately to the rainfall occurrence at each of the stations in Zambia and Ghana for the gauge observations as well as the REs.

The seasonality at the Zambian stations were well captured by all the REs, with the highest rain day frequency in the November to May season (see Figures~\ref{fig_16}-\ref{fig_19} for ERA5, TAMSAT, CHIRPS, and ENACTS respectively consistent with the REs not shown here). 

At the 0.85 mm rain day threshold, the REs overestimated the rain day frequency in varying degrees. In general, ENACTS seemed to have the lowest overestimation, followed by CHIRPS and TAMSAT. ERA5 (and AGERA5, PCDR, CHIRP, not shown here) had a higher degree of overestimation. 

Most of the REs tended to match the gauge observations better at a higher rain day frequency, while still capturing well the seasonality.   

There was an indication that the choice of threshold may depend on the the time of year. For example, the choice of 5 mm for ERA5 rain day threshold seemed to be optimal for most parts of the year at Mpika, but was not optimal between January and March (see Figure \ref{fig_16}). 

There was also an indication that the choice of threshold may depend on the location. For instance, the choice of 5 mm rain day threshold seemed optimal for ERA5 at Solwezi, but not at Zambezi (see Figure \ref{fig_16}). 

Last but not least, the choice of threshold also appeared to be dependent on the product. For example, while the choice of 5 mm rain day threshold appeared to be optimal for TAMSAT at Magoye (Figure \ref{fig_17}), it was not optimal for ENACTS at the same station (Figure \ref{fig_19}).

Similar to the case of Zambia, all the REs captured the seasonality well at the southern stations of Ghana, capturing the bi-modal rainfall patterns (April-July, and September-November, with peaks around June and October respectively).

However, at 0.85mm threshold, CHIRPS overestimated rain day frequency between February and June while it underestimated rain day frequency between June and October at Abetifi, Akim Oda, Akuse, Axim, Ho, KIAMO Accra, Koforidua, Kumasi, Saltpond, Sefwi Bekwai, Sunyani, and Takoradi (Figure \ref{fig_22}). TAMSAT also underestimated rainday frequency between June and October at Akim Oda, Axim, Koforidua, Sefwi Bekwai, and Takoradi (Figure \ref{fig_21}). While increasing thresholds seemed to fix the overestimation for the other REs across all stations, the underestimation by CHIRPS and TAMSAT did not seem to change under different rain day thresholds. There is potentially an issue with these REs at these locations.  

Aside from CHIRPS and PCDR, the REs also captured seasonality of the northern stations of Ghana well, capturing the unimodal patterns. CHIRPS and PCDR (Figures~\ref{fig_22} and \ref{fig_20} respectively) exhibited bi-modal patterns at these northern stations, which have unimodal rainfall patterns.  

Similar to Zambia, most of the REs tended to match the gauge observations better at a higher rain day frequency, while still capturing well the seasonality. These thresholds seemed to depend on the RE, location and time of year (similar to the case of Zambia).

\subsection{Rainfall Intensity}
\begin{figure}[ht]%
	\centering
	\includegraphics[width=0.5\textwidth]{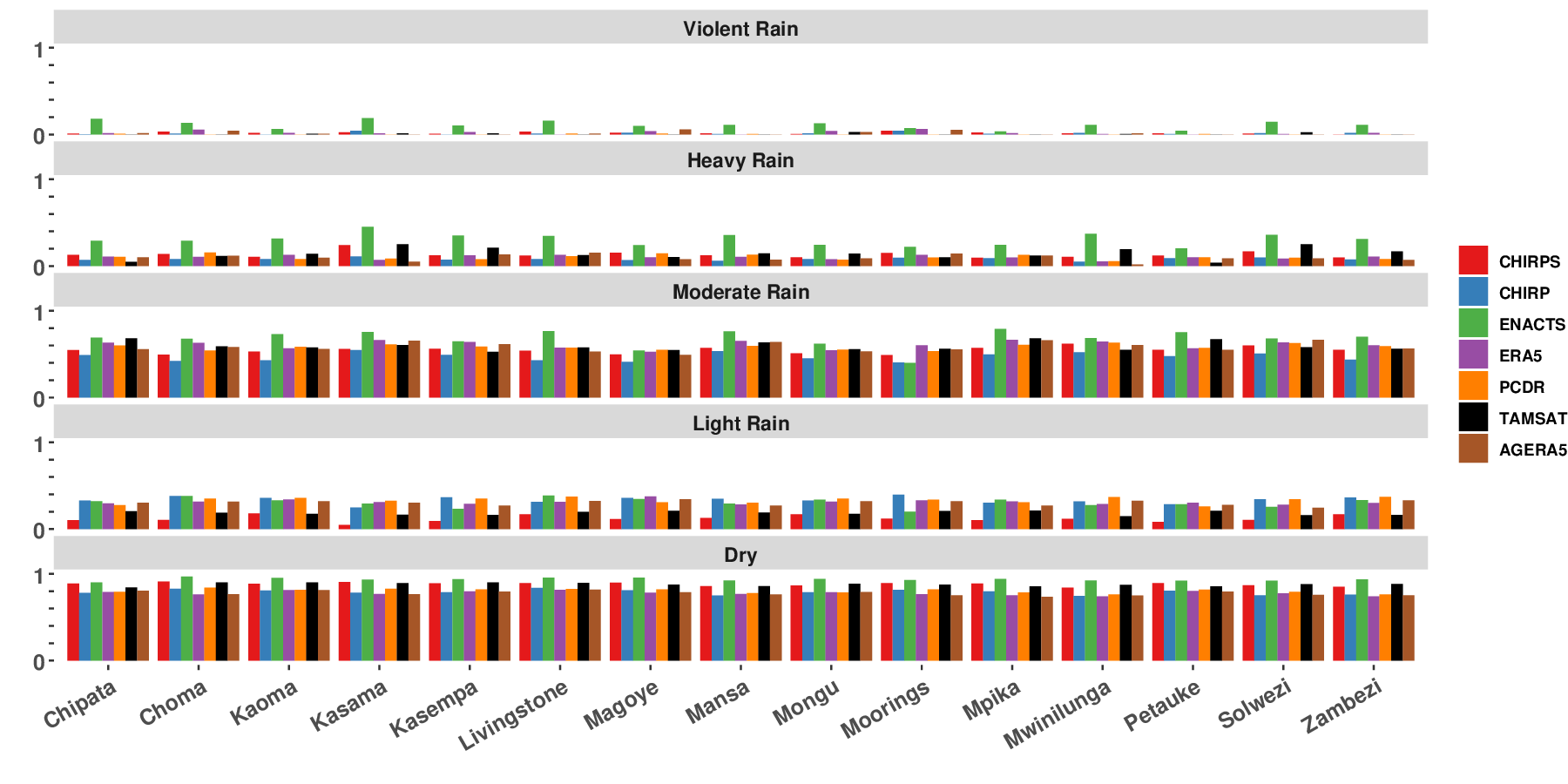}
	\caption{POD for different rainfall intensity categories in Zambia}\label{fig_23}
\end{figure}

\begin{figure}[ht]%
	\centering
	\includegraphics[width=0.5\textwidth]{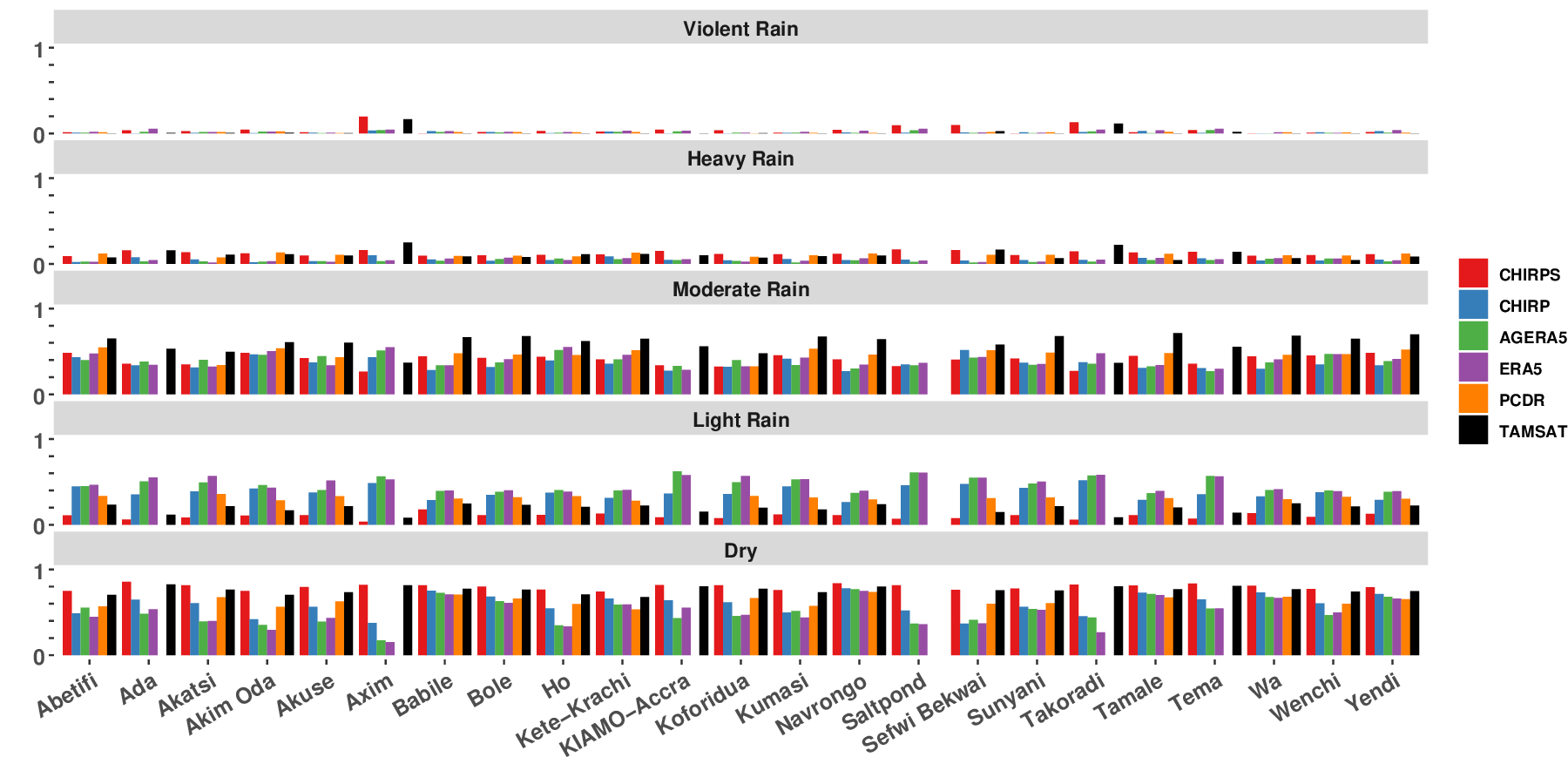}
	\caption{POD for different rainfall intensity categories in Ghana}\label{fig_24}
\end{figure}

Figure \ref{fig_23} and \ref{fig_24} show the different REs and their POD for different rainfall intensity categories at the different stations in Zambia and Ghana respectively. 

\subsubsection{Dry days detection} 

All the REs showed high skill of detecting dry days (with 70\% $<$ POD $<$ 100\%) across all stations in Zambia (Figure~\ref{fig_23}). ENACTS, CHIRPS and TAMSAT had about 10\% chance higher in detecting dry days than the other REs. In the case of Ghana, not all the REs demonstrated a high probability of accurately detecting dry days, as shown in Figure \ref{fig_24}. Aside from TAMSAT and CHIRPS, which exhibited a POD of at least 70\% across all stations, the other REs had POD often below 70\% in southern Ghana, and  60\% $<$ POD $<$ 85\% in northern Ghana. The worst performance of most of the REs in detecting dry days was observed in stations located in the central and southern regions, particularly along the southwestern coast of the country. 

\subsubsection{Light and moderate rains detection}
In the case of Zambia all the REs (with the exception of CHIRP) had POD $>$ 50\% in detecting moderate rain at almost all the stations, with ENACTS having 50\% $<$ POD $<$ 80\%, outperforming all the REs, followed by TAMSAT (with 50\% $<$ POD $<$ 70\%). All the REs had a probability of less than 50\% in detecting light rain.

In the case of Ghana, the two reanalysis products --- ERA5 and AGERA5 --- stood out slightly in the detection of light rain, even though they both had a POD below 50\% at most of the stations. TAMSAT outperformed all the REs in the detection of moderate rain (with 50\% $<$ POD $<$ 80\%) at almost all the stations.

\subsubsection{Heavy and violent rains detection}
All the REs all performed poorly in detecting heavy and violent rains (with POD close to zero across all stations) in both Ghana and Zambia (see Figure~\ref{fig_23} and \ref{fig_24}). 

\begin{figure*}[h!]
	\centering
	\begin{minipage}{\textwidth}
		\centering
		\includegraphics[width=\textwidth]{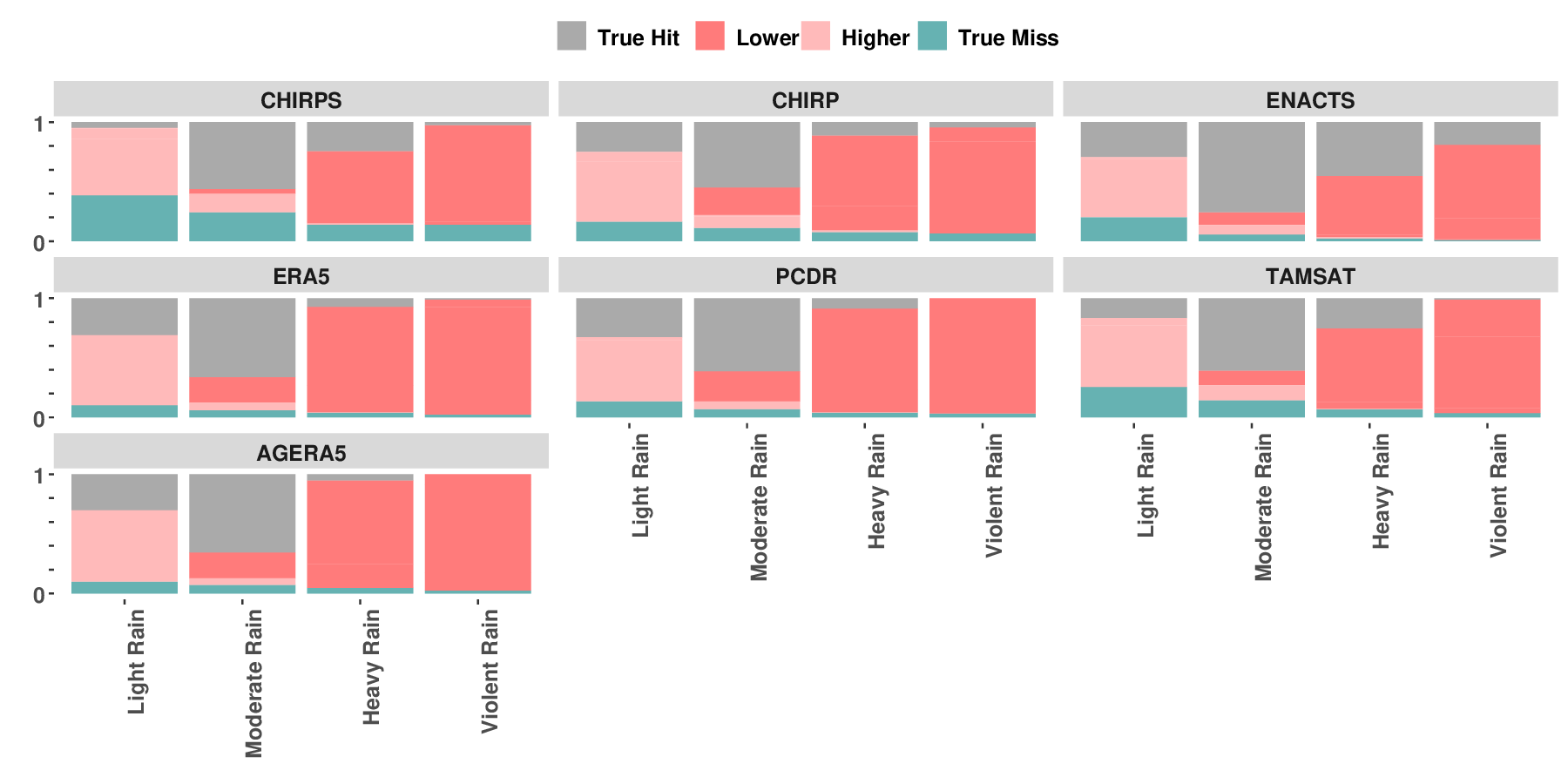}
		\caption{Graph of Kasama station rainfall intensity categories. True Hit represents the proportion of rainfall events in the rainfall intensity category that were correctly detected by the RE; True Miss represents the proportion of rainfall events in the rainfall intensity category that the RE detected as a dry day. Lower represents the proportion of rainfall events in the rainfall intensity category that the RE missed by estimating a lower intensity category. Higher represents the proportion of rainfall events in the rainfall intensity category that the RE missed by estimating a higher intensity category.}\label{fig_25}
	\end{minipage}
	
	\vspace{0.5cm} 
	
	\begin{minipage}{\textwidth}
		\centering
		\captionof{table}{Percentages of observed rainfall intensity categories at the various stations in Zambia}\label{tab_nrain_cat}%
		\begin{tabular}{@{}lllll@{}}
			\toprule
			Station & Light Rain (\%) & Moderate Rain (\%) & Heavy Rain (\%) & Violent Rain (\%) \\
			\midrule
			Chipata & 36  & 39    & 16  & 9 \\
			Choma   & 40   & 43    & 12   & 5 \\
			Kaoma   & 42 & 41   & 13  & 5 \\
			Kasama  & 37   & 42    & 16   & 5 \\
			Kasempa & 38  & 42   & 15   & 5 \\
			Livingstone & 43  & 40    & 12   & 5 \\
			Magoye  & 41   & 40    & 14  & 5  \\
			Mansa   & 37  & 43    & 15  & 6  \\
			Mongu   & 39  & 42    & 14   & 5  \\
			Moorings & 34   & 46   & 14   & 6  \\
			Mpika   & 41  & 40    & 15   & 5  \\
			Mwinilunga & 41   & 41    & 13   & 4  \\
			Petauke & 39  & 39    & 16   & 6 \\
			Solwezi & 36   & 43    & 16   & 5  \\
			Zambezi & 40  & 42   & 14   & 5  \\
			\botrule
		\end{tabular}
	\end{minipage}
\end{figure*}  

Figure \ref{fig_25} represents the distribution of hits and misses in the different rainfall intensity categories at Kasama, Zambia. The distribution of hits and misses in Kasama was not very different from those at other stations in Zambia and Ghana (not shown). True Hit was relatively high for moderate rain, which relates to the observation in Figure \ref{fig_23}. As expected, True Miss was very low across all intensity categories. True Miss was the lowest under violent rains followed by heavy rain. The figure further shows that rainfall was correctly detected by most of the REs on days of violent and heavy rains, but the intensity was usually rather underestimated. 

Table \ref{tab_nrain_cat} also gives percentages of different rainfall intensity category observed at the various stations in Zambia. This table shows that heavy and violent rains were usually less than 20\% and 10\% respectively of all observed rainfall rainfall events across all stations.

These results highlight the strengths and limitations of REs, with implications for their application in climate studies such as drought, and extreme event analysis. The results are discussed in the next section.

\section{Discussion}\label{section_discussion} 
The results have shown that no single product is universally optimal across all contexts highlighting the fact that product recommendations must be specific to the intended application, addressing questions such as:
\begin{enumerate}
	\item Can the product reliably monitor heavy and violent rain events in region X? 
	
	\item Is it suitable for drought analysis in region Y?
	
	\item Does the product adequately capture seasonality in location Z?
\end{enumerate}This targeted validation enables more informed and context-specific decision-making. For instance, improved validation of REs could help farmers optimize irrigation strategies, reduce crop losses due to drought.

The topography in the southern part of Ghana (including locations like Axim and Takoradi) is complex due to the vegetation and terrain, and its proximity to the coast and surrounding forest regions. These regions are subject to localized weather patterns \citep{Amekudzi2015-qk}, which can be harder for the REs to accurately capture. Coastal influences, such as sea breezes, can induce small-scale convective systems that could be missed by the REs. On the contrary, Zambia is predominantly flat and landlocked, with rainfall largely driven by convective systems, and there is typically a strong correlation between the cold cloud tops of these systems and rainfall \citep{Maidment2017}. Our results confirm the fact performance of the REs under annual summaries, seasonality studies, and rainfall intensity detection was more consistent in Zambia as compared to Ghana.

Previous works have shown that REs are usually good at detecting dry days at many places \citep{Mekonnen2023, ZambranoBigiarini2017}. From our results, all the REs were seen to detect dry days well in Zambia and northern Ghana while CHIRPS and TAMSAT still maintained consistent high performance in south Ghana as well, consistent with previous existing literature. One explanation could be as explained in the previous paragraph. A reason for the superior performance of CHIRPS and TAMSAT on detecting dry days could be due the fact that they have been designed for drought monitoring \citep{Funk2015, Maidment2017}. 

It is known that the REs are extensively validated on their ability to capture daily, sub-seasonal, seasonal, or annual rainfall amounts during development \citep{Funk2015, Maidment2017, Ashouri2015}. Our results shown on Figures~\ref{fig_12} and \ref{fig_13} show low biases and good correlations on total rainfall, consistent with \citep{Ageet2022}, and likely explained by the literature above.

All the REs capture the seasonality well in both countries at the 0.85mm (or higher, or both) rain day threshold. The good performance of the REs at the seasonal scale is in agreement with results by other researchers \citep{Ageet2022}. 

ENACTS includes a large number of meteorological station data than other similar blended REs \citep{siebert2019evaluation, Dinku2022}. CHIRPS also includes some station data, while TAMSAT is calibrated with station data. Our results showed that ENACTS stood out as one of the best products across multiple contexts in Zambia (followed by CHIRPS and TAMSAT). This is likely due to the merging/calibration of these products with station data. Local calibration could be
a potential approach for reducing the satellite precipitation errors \citep{hess-21-6201-2017, Dinku2008, Dinku2011}. 

REs have been known with their tendency to overestimate the number of rainy days, as confirmed by previous works within the Africa region \citep{MAPHUGWI2024107718}. Our results as shown in Figures~\ref{fig_8} and \ref{fig_9} are consistent with existing findings. This tendency to overestimate rainy days is expected, as the REs represent area-based precipitation estimates, while gauge observations are point measurements. Due to the overestimation, they had low instances of False Negatives. This is confirmed by the results in Figures \ref{fig_10} and  \ref{fig_11} for Zambia and Ghana respectively. They also had considerable proportions of False Positives due to the overestimation (see Figures~\ref{fig_10} and  \ref{fig_11}). Some of these False Positives may have stemmed from rainfall occurring elsewhere within the grid area but outside the immediate vicinity of the gauge. Further investigation, using additional station data from within these grids, could clarify these discrepancies. It is also possible that rainfall detected by REs occurs several kilometers away or even on a different day, as noted by \cite{Houze2004}. Rain-bearing clouds can travel hundreds of kilometers before releasing rain, sometimes accumulating over several days before delivering intense rainfall in a single event \citep{Guilloteau2016}. 

All REs exhibited poor performance in detecting heavy and violent rains (see Figures \ref{fig_23} and \ref{fig_24}), consistent with findings reported in existing literature \citep{Yang2016, Mekonnen2023, ZambranoBigiarini2017, Ageet2022}. This suggests that, in their current state, these products are likely unsuitable for detecting heavy and violent rains or for conducting related studies in these regions. The challenges in accurately capturing these events may be attributed to spatial resolution constraints, limitations in rainfall retrieval algorithms, and regional variability \citep{hess-21-6201-2017, Mekonnen2023}. Our results are based on a point-to-pixel validation approach, where spatially averaged RE values from their native pixels are directly compared with point-based station observations \citep{Maranan2020}. This approach may dilute the intensity of localized heavy and violent rains, leading to underestimation. Furthermore, the rainfall variability in our study regions, which is predominantly driven by convective systems, poses additional challenges for REs in detecting the heavy and violent rains associated with these systems \citep{hess-21-6201-2017}. 

Despite the REs' limitation in detecting heavy and violent rains, Figure \ref{fig_25} demonstrates that most days of heavy and violent rains were correctly classified as rainy by the REs, albeit with underestimated rainfall amounts. This indicates that the products have the potential to detect heavy and violent rains but require targeted improvements, such as region-specific calibration, to better capture the intensity and distribution of heavy rainfall \citep{Fang2019}. Developing REs specifically tailored to heavy and violent rains detection, which could complement existing products, is essential for enhancing climate services in regions vulnerable to these events.

\section{Conclusion}\label{sec_conclusion}
This study evaluated the performance of eight REs --- CHIRPS, TAMSAT, CHIRP, ENACTS, ERA5, AgERA5, PERSIANN-CDR, and PERSIANN-CCS-CDR --- in Zambia and Ghana using a point-to-pixel validation approach. The analysis encompassed spatial consistency, annual rainfall summaries, seasonal patterns, and rainfall intensity detection across 38 ground stations. The results have provided useful insights into the performance of the products and the implication on their usage as follows: 

All products exhibited high POD for dry days in Zambia and nothern Ghana (with 70\% $<$ POD $<$ 100\%, and 60\% $<$ POD $<$ 85\% respectively) while CHIRPS and TAMSAT maintained a consistent high performance in southern Ghana. These products may potentially be useful for drought detection at those locations.

There were biases in the products under multiple summaries (including total rainfall, rain day frequency). Bias-correction is recommended, especially for reducing the biases on the number of rainy days. Numerous studies have demonstrated effective bias-correction techniques for REs (e.g., \citep{Gudmundsson2012, Schmidli2006}), which could be adapted for this purpose. However, the application of bias-correction methods and the recommendation of specific techniques for our study regions were beyond the scope of this study.

The products merged/calibrated with station data (ENACTS, CHIRPS, and TAMSAT) appeared to perform better than the other REs under many of the contexts. This is perhaps one potential route for improving the performance of REs. 

The results from this study are likely applicable to regions with similar climatic characteristics, particularly in sub-Saharan Africa and other areas reliant on rain-fed agriculture. However, we recommend validation in such locations according to the use case. The methods used in this work can be readily applied to new locations, even at one station. This makes this approach particularly valuable for regions facing similar climatic and agricultural challenges, where reliable rainfall data are critical for climate services and decision-making.	  
	
Lastly, all REs performed poorly in the detection of heavy and violent rains (with POD close to 0\%). At their current state, the products may not be recommended for heavy and violent rains detection at these locations, such as floods. Future research should look more into this.

\backmatter
 
\bmhead{Supplementary information}

Additional figures and tables are in the Supplementary Material (Tables E1-E6, and Figures E1-E69).

\bmhead{Acknowledgments}

The authors acknowledge and thank the Ghana Meteorological Department, the Zambia Meteorological Department for their collaboration, without which this study would not have been possible.

\section*{Declarations}

\begin{itemize}
\item \textbf{Funding} This publication was made possible by a grant from Carnegie Corporation of New York (provided through the African Institute for Mathematical Sciences). The statements made and views expressed are solely the responsibility of the author(s).

\item \textbf{Competing interests} The authors have no relevant financial or non-financial interests to disclose.

\item \textbf{Ethics approval} Not applicable 
\item \textbf {Consent to participate} Not applicable

\item \textbf{Consent for publication} Not applicable

\item \textbf{Data Availability} The RE data used are all publicly available. The station data for Ghana and Zambia can be obtained from GMet and ZMD respectively. 

\item \textbf{Code availability} Code is available upon a reasonable request.

\item \textbf{Authors' contributions} 
Conceptualization: J.B., D.S.; Methodology: J.B., D.S.; Formal analysis: J.B., D.S.; Data curation: J.B., D.S., F.F.T.; Writing --- original draft: J.B.; Writing --- review \& editing: J.B., D.S., D.N., F.F.T.; Visualization: J.B., D.S.; Supervision: D.S., D.N., F.F.T.

\end{itemize}

\noindent

\bigskip

\end{document}